\documentclass[aps,twocolumn,superscriptaddress]{revtex4-1}

\usepackage[a4paper,left=2.0cm,right=2.2cm,top=2.0cm,bottom=2.0cm]{geometry}
\usepackage{graphicx}
\usepackage{amsmath}
\usepackage{amssymb}
\usepackage{color}
\usepackage{ulem,bbm}
\usepackage{paralist,hyperref} 
\usepackage[small,compact]{titlesec}

\titlespacing{\section}{0ex}{2ex}{1ex}
\titlespacing{\subsection}{0ex}{1ex}{0.5ex}

\def\be{\begin{eqnarray}}
\def\ee{\end{eqnarray}}

\def\d{\!\!\!\mathrm{d}}
\def\D{\mathrm{d}}
\def\tr{\mathrm{tr}}
\def\Green{G}

\newcommand\ind[1]{^{(#1)}} 
\newcommand\indd[1]{^{(#1) \, \dag}} 
\newcommand\indT[1]{^{(#1) \, \rm T}} 
 
\def\H{\hat{H}}
\def\HS{\hat{H}_{S}}
\def\HR{\hat{H}_{R}}
\def\Vint{\hat{V}_{\rm int}}
\def\Op{\hat{O}}

\newcommand{\Pwwo}{\hat{\boldsymbol{\Pi}}_{\omega}}

\newcommand{\Xwwo}{\hat{\boldsymbol{X}}_{\omega}}

\newcommand{\Spinwo}{\hat{\boldsymbol{S}}}

\newcommand{\Spin}[1]{\Spinwo\ind{#1}\!}
\newcommand{\Pw}[1]{\Pwwo\ind{#1}\!}
\newcommand{\Xw}[1]{\Xwwo\ind{#1}\!}
\newcommand{\aw}[1]{\hat{\boldsymbol{a}}_{\omega}\ind{#1}\!}
\newcommand{\awd}[1]{\hat{\boldsymbol{a}}_{\omega}\indd{#1}\!}

\newcommand{\Scomp}[2]{\hat{S}\ind{#1}_{#2}}
\newcommand{\Xwcomp}[2]{{\hat{X}_{\omega, #2}}\ind{#1}}
\newcommand{\Pwpcomp}[2]{{\hat{\Pi}_{\omega', #2}}\ind{#1}}

\newcommand{\awcomp}[2]{\hat{a}_{\omega,  #2}\ind{#1}}

\newcommand{\awpdcomp}[2]{\hat{a}_{\omega', #2}\indd{#1}}

\def\Bext{\boldsymbol{B}_{\rm ext}}
\newcommand{\Benv}[1]{\hat{\boldsymbol{B}}_{\rm env}\ind{#1}}
\newcommand{\bstoch}[1]{\hat{\boldsymbol{b}}\ind{#1}\!}
\newcommand{\bstochwo}{\hat{\boldsymbol{b}}}
\newcommand{\bstochcomp}[2]{\hat{b}_{#2}\ind{#1}\!}
\newcommand{\bstochdcomp}[2]{\hat{b}_{#2}\indd{#1}}
\newcommand{\Beff}[1]{\hat{\boldsymbol{B}}_{\rm eff}\ind{#1}\!}
\newcommand{\Beffwo}{\hat{\boldsymbol{B}}_{\rm eff}}

\newcommand{\kernelwo}{{\mathcal K}}
\newcommand{\kernel}[1]{\kernelwo\ind{#1}\!}
\newcommand{\tkernelwo}{\tilde{\kernelwo}}
\newcommand{\tkernel}[1]{\tkernelwo\ind{#1}}

\newcommand{\PSD}{\mathcal{P}}

\newcommand{\clspin}{\boldsymbol{S}}

\newcommand{\tin}{t_0} 

\newcommand{\Tesla}{~\mbox{T}}

\renewcommand{\sec}{~\mbox{s}}
\newcommand{\Kelvin}{~\mbox{K}}

\def\Jnm{\mathcal{J}\ind{nm}}

\newcommand{\Cwtens}{{\mathcal C}_{\omega}}
\newcommand{\Cw}[1]{{\mathcal C}_{\omega}\ind{#1}}
\newcommand{\CwT}[1]{{\mathcal C}_{\omega}\indT{#1}}

\newcommand{\Cwwo}{ C_{\omega}}

\newcommand{\pow}{\tilde{p}}

\newcommand{\Lor}{{\mbox{\sf \scriptsize Lor}}}

\newcommand{\Ohm}{{\mbox{\sf \scriptsize Ohm}}}
\newcommand{\class}{{\mbox{\sf \scriptsize cl}}}
\newcommand{\quant}{{\mbox{\sf \scriptsize qu}}}
\newcommand{\stph}{{\mbox{\sf \scriptsize stat phys}}}

\newcommand{\Appendix}{Appendix} 

\begin{document}

\title{Quantum Brownian Motion for Magnets}

\author{J. Anders}
\email[]{janet@qipc.org}
   \affiliation{Department of Physics and Astronomy, University of Exeter, Stocker Road, Exeter EX4 4QL, UK.}
    \affiliation{Institut f\"ur Physik und Astronomie, University of Potsdam, 14476 Potsdam, Germany.}
 
\author{C.R.J. Sait}
   \affiliation{Department of Physics and Astronomy, University of Exeter, Stocker Road, Exeter EX4 4QL, UK.}

\author{S.A.R. Horsley}
   \affiliation{Department of Physics and Astronomy, University of Exeter, Stocker Road, Exeter EX4 4QL, UK.}


\begin{abstract}

Spin precession in magnetic materials is commonly modelled with the classical phenomenological Landau-Lifshitz-Gilbert (LLG) equation.  Based on a quantized spin+environment Hamiltonian, we here derive a general spin operator equation of motion that describes three-dimensional precession and damping and consistently accounts for effects arising from memory, coloured noise and quantum statistics. The LLG equation is recovered as its classical, Ohmic approximation. 
We further introduce resonant Lorentzian system--reservoir couplings that allow a systematic comparison of dynamics between Ohmic and non--Ohmic regimes. 
Finally, we simulate the full non-Markovian dynamics of a spin in the semi--classical limit. At low temperatures, our numerical results demonstrate a characteristic reduction and flattening of the steady state spin alignment with an external field, caused by the quantum statistics of the environment. The  results provide a powerful framework to explore general three-dimensional dissipation in quantum thermodynamics.

\end{abstract}

\maketitle

The continued miniaturisation of critical components in consumer electronics and neighbouring technologies will require  a deeper understanding of thermal noise and general thermodynamic principles beyond the classical macroscopic world. Quantum thermodynamics   \cite{Goold2016,Vinjanampathy2016,Book2018} has emerged as a field addressing the conceptual challenges related to the exchange of energy and information at the nanoscale. 
Recent advances include  studies of heat transport in quantum systems~\cite{Wichterich2007,Boudjada2014,Yang2014,Freitas2017,Funo2018,Whitney2018,Yang2019,Benatti2020}, the characterisation of memory effects in their dynamics~\cite{Maniscalco,Rivas2010,fermionic-Chen2013,fermionic-Strasberg2016,deVega,Cianciaruso2017,Strasberg2018,Raja2018},  and clarification of the impact of quantum coherence and correlation on thermodynamic processes~\cite{Uzdin2015,Brask2015,Kammerlander2016,Sapienza2019,Klatzow2019}. 
The establishment of a generalised thermodynamic framework, valid for nanoscale systems that strongly couple to environmental modes, is well under way~\cite{Seifert2016, philbinanders2016, Jarzynski2017, Miller2017,Cresser2017,Miller2019,Kawai2019,Strasberg2020,Purkayastha2020}, and for magnetic molecules an environment-induced renormalisation of the anisotropy has been predicted~\cite{Kenawy2018}. 
Two open quantum systems models have served as the workhorse for many of these conceptual studies; the Caldeira-Leggett model for quantum Brownian motion \cite{caldeira,Hu92,Funo2018} and the spin-boson model of a spin (or many spins) coupled to a one-dimensional harmonic bath \cite{Thoss2001,Breuer-Petruccione,ABV2007,Boudjada2014,Yang2014,Purkayastha2020}. 
These describe a very wide range of physical situations extending to studies of quantum effects in bio-chemical reactions \cite{Huelga2013}, where they are used to model exciton-phonon interactions~\cite{Nazir2016}.  

Until now few nanoscale technologies have required the use of advanced open quantum systems techniques. 
But advances in engineering magnetic materials for magnetic hard drives at unprecedented length and time-scales \cite{Seagate} are likely to require a more detailed picture of spin dynamics including memory and quantum signatures.  
Here we introduce a three-dimensional open quantum system model to characterise the quantum Brownian motion of spins in magnetic materials.

Magnetic behaviour has been studied extensively based on the classical phenomenological Landau--Lifshitz--Gilbert (LLG) equation~\cite{gilbert,mayergoyz2009,lakshmanan2011,vansteenkiste2014,evans2014} 
\begin{equation}
	\frac{\partial\boldsymbol{M}}{\partial t}=\gamma\boldsymbol{M}\times\left[\boldsymbol{B}_{\rm eff}-\eta_G\frac{\partial\boldsymbol{M}}{\partial t}\right], \label{eq:llg_equation}
\end{equation}
which is routinely solved with micromagnetic and atomistic simulations.
Here $\boldsymbol{M}$ is the magnetic moment, $\gamma$ is the gyromagnetic ratio and $\boldsymbol{B}_{\rm eff}$ is an effective magnetic field which includes the external field~
\footnote{Note that Eq.~\eqref{eq:llg_equation} is expressed in  SI rather than Gaussian units 
(as in e.g.~\cite{lakshmanan2011}).}, 
exchange and anisotropy effects, as well as stochastic magnetic noise $\boldsymbol{b} \propto \sqrt{T}$ stemming from an environment at temperature $T$ that was added by Brown~\footnote{Not the Brownian motion Brown!} in 1963 \cite{brown1963}. 
The final term on the right of~\eqref{eq:llg_equation} is the so--called ``Gilbert damping'' term and the positive constant $\eta_G$ is the damping parameter~
\footnote{Using vector identities the time derivative on the right hand side of Eq.~\eqref{eq:llg_equation}  can be eliminated and the equation becomes $\partial\boldsymbol{M} / \partial t=\gamma'\boldsymbol{M}\times \boldsymbol{B}_{\rm eff}-\lambda \boldsymbol{M}\times \left(\boldsymbol{M}\times  \boldsymbol{B}_{\rm eff} \right)$ with $\gamma'$ and $\lambda$ functions of $\gamma, \eta_G$ and $|\boldsymbol{M}|$.}, which is often rewritten as $\eta_{G}=\eta/|\boldsymbol{M}||\gamma|$ with a unit-free $\eta$. 

Gilbert damping is not derived from microscopic principles, but chosen as the simplest term that could serve to align the magnetic moment with the applied field~\cite{gilbert}.
As we will see, it contains no memory which is increasingly seen as a limitation~\cite{ciornei2011,Neeraj2021}.  
Advances in engineering magnetic materials at the nanoscale and manipulating them on ultrafast timescales indicate that a theory beyond the classical LLG equation is required~\cite{evans2014}. 
Early attempts 
have pursued a path integral derivation of a quantum spin dynamics equation  \cite{rebei2003}, as well as other conceptually related classical and quantum derivations~\cite{garcia-palacios1999,Rossi2005}. These derivations were not directly applied to the calculation of magnetization dynamics or steady states, nor have they been connected to recent generalizations of Gilbert damping that include inertial terms \cite{ciornei2011,Bauer08,bose2011,schutte2014,thonig2015,bajpai2019,li2015,Neeraj2021} or provided an assessment of quantum effects.

Here we go further and develop a comprehensive and quantum-thermodynamically consistent theory suitable to describe the quantum dynamics of spins in magnetic materials including  non--Markovian damping, coloured noise and quantum zero-point fluctuations. 
%
Unlike the conceptually pioneering Caldeira-Leggett model that has few experimental realisations, the developed three-dimensional quantum spin model is directly applicable for atomistic spin dynamics simulations~\cite{evans2014,Barker2019}, ultrafast magnetism experiments~\cite{beaurepaire1996}, and systems exhibiting anisotropic damping~\cite{Chen2018}.

The paper is organised as follows: In section  \ref{sec:equationderivation} the general quantum spin dynamics equation for spin operator precession in three dimensions is derived. For the simplest, Ohmic, coupling this equation is found to reduce to the memory-free LLG equation. In section \ref{sec:couplings} we introduce Lorentzian couplings as a systematic method for exploring non-Markovian dynamical regimes in general open quantum systems, including spins. Finally, in section \ref{sec:simulations} we detail a numerical method to simulate non-Markovian dynamics, and  present results for a single classical spin that illustrate the differences between spin dynamics and steady states arising with non-trivial memory, coloured noise, and quantum bath statistics in comparison to those obtained with the memory-free LLG equation.  Conclusions and open questions are discussed in section \ref{sec:conclusion}.

\section{\bf Quantum spin dynamics equation} \label{sec:equationderivation}

\subsection{\bf System+environment Hamiltonian}

We begin by introducing the quantized Hamiltonian describing the different contributions to the total energy of the system, consisting of spins as well as environmental degrees of freedom (e.g. electrons and phonons), given by
\be
	\H = \HS + \HR + \Vint , 
	\label{eq:total_hamiltonian}
\ee
where $\HS$ is the bare spin Hamiltonian operator which captures the spin energy in external fields and interactions between spins,  $\HR$ is the environmental or {\it reservoir} Hamiltonian, and $\Vint$ is the interaction between the spins and the reservoir.

We choose $\HS$ as the sum of the interaction with a homogeneous external field $\Bext$~\footnote{We use SI rather than Gaussian units, so we have $B$ (units ${\rm mass}/({\rm charge}\times{\rm time})$) rather than an $H$-field (units ${\rm charge}/({\rm length}\times{\rm time})$).} and the exchange interaction between three-dimensional spin vector operators $\Spin{n} =(\Scomp{n}{1},\Scomp{n}{2},\Scomp{n}{3}$) at sites $n$ of a lattice~\footnote{Here the spins are discrete and positioned on a lattice, but one could also use a continuum description, as in micromagnetics~\cite{mayergoyz2009}.},
\be  \label{eq:HS}
	\HS= 
	- \gamma \sum_{n} \Spin{n}  \cdot \Bext  
	-\frac{1}{2}\sum_{n, m\neq n} \Spin{n}  \cdot \Jnm   \Spin{m} . \quad \,\,
\ee
Here $\Jnm$ is the exchange tensor for spin pairs $(n,m)$ \footnote{Note that tensors and vectors are set in calligraphic and bold font, respectively, and that scalar products between vectors are indicated with $\cdot$, while a tensor followed by a tensor or a vector is to be understood as matrix multiplication.},
which can include the Dzyaloshinskii-Moriya interaction~\cite{evans2014}. It is straightforward to include additional energetic terms in the bare spin Hamiltonian, such as energies associated with magnetic anisotropy.
Instead of the magnetic moment $\boldsymbol{M}$ used in  Eq.~\eqref{eq:llg_equation}, we will here work with the spin angular momentum $\boldsymbol{S}$ proportional to $\boldsymbol{M}$, $\boldsymbol{M}=\gamma\boldsymbol{S}$, where $\gamma$ is the gyromagnetic ratio. In the following we will assume the gyromagnetic ratio $\gamma=-g_{e}\mu_{B}/\hbar = - 1.76 \cdot 10^{11} \sec^{-1} \Tesla^{-1}$ for an electron. 

\begin{figure}[b]
\includegraphics[width=0.43\textwidth]{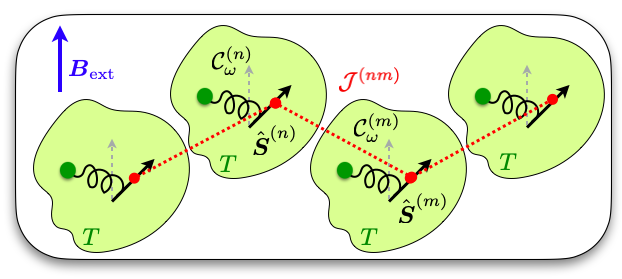}
\caption{\label{fig:model} \sf {\it Illustration of Hamiltonian model.} In addition to precessing in an external field $\Bext$, and coupling to its spin neighbours $m$ with strength $\Jnm$, each spin $\Spin{n}$ couples to its environmental mode (phonons, electrons) at frequency $\omega$ with a coupling function $\Cw{n}$. All environmental modes are assumed to be thermal at the same temperature $T$.}
\end{figure}

The reservoir Hamiltonian is commonly modelled as a set of harmonic oscillators \cite{caldeira,huttner1992}, and we here follow the continuous reservoir approach by Huttner and Barnett \cite{huttner1992}, taking the reservoir Hamiltonian as
\be
	\HR =\frac{1}{2} \sum_{n} \int_{0}^{\infty} \d\omega 
	\left[ \left(\Pw{n}\right)^2 + \omega^{2}  \left(\Xw{n}\right)^2 \right] . \label{eq:HR}
\ee
It describes a continuous frequency reservoir at each lattice site $n$, where $\Pw{n}$ and $\Xw{n}$ are (three-dimensional) momentum and position operators of the reservoir oscillator with frequency $\omega$.
The position operators $\Xw{n}$ physically represent variations in the environment to which the spin at site $n$  responds, see illustration Fig.~\ref{fig:model}, as for example, in magnon--phonon mediated loss~\cite{azzawi2017}. 
Unlike most  system+environment Hamiltonians which assume one-dimensional coupling, we here take the spin-reservoir interaction to be of the three-dimensional form
\be  \label{eq:Vint}
	\Vint	= - \gamma \sum_{n} \Spin{n} \cdot \int_{0}^{\infty} \d\omega \, \,\Cw{n}  \Xw{n}.
\ee
This coupling allows angular momentum transfer as well as energy transfer between the spins and the environment. Here $\Cw{n}$ is a three-dimensional coupling tensor and a function of frequency $\omega$. At each $\omega$, the coupling tensor determines the strength of the coupling of each spin to its reservoir oscillators at frequency $\omega$, thus acting as a frequency filter. As we shall see, the choice of the coupling $\Cw{n}$ will determine the damping of the spin dynamics as well as the stochastic noise experienced by the spins.  

For readers concerned about time-reversal symmetry of $\Vint$ in \eqref{eq:Vint}, we note that $\Xw{n}$ should be interpreted as an effective magnetic field seen by the spins due to their interaction with the environment, which has the same time symmetry as $\Spin{n}$ \footnote{Note that in contrast to what is typically done in Caldeira--Leggett type models\cite{caldeira} no counter term has been included here. In any case, coupling to a spin would result in a term proportional to $\boldsymbol{S}^2 \propto \mathbbm{1}$, which will incur an offset in the overall Hamiltonian that does not affect the dynamics.}.
Indeed, the theory of magnetic materials developed here on the basis of the system+environment Hamiltonian $\H$ is analogous to macroscopic QED, an effective medium theory which successfully describes quantum electromagnetism in dielectric materials~\cite{huttner1992,scheel2008,philbin2010}.  Instead of trying to give a fully microscopic description that accounts for every light-matter interaction in the material, macroscopic QED characterizes electromagnetic materials in terms of two measured frequency dependent susceptibilities.  The quantum Hamiltonian is then written in terms of these susceptibilities, and can be used in applications from predicting the Lamb shift to the Casimir effect~\cite{scheel2008,philbin2011}.

\subsection{\bf Equations of motion}

Having set up the full Hamiltonian~\eqref{eq:total_hamiltonian} of the spins and environment degrees of freedom  allows the study of the spins' reduced state dynamics  $\hat{\rho}_S (t) = \tr_R\left[ \hat{\rho}_{SR} (t)  \right]$ of the total state
\be
    \hat{\rho}_{SR} (t) = e^{{\rm i} \H t/\hbar} \, \left( \hat{\rho}_S(0) \otimes \hat{\rho}_R \right) \, e^{-{\rm i} \H t/\hbar},
\ee
where the  reservoir state $\hat{\rho}_R = e^{-\beta \HR}/\tr[e^{-\beta \HR}]$ is thermal at some inverse temperature $\beta=1/k_B T$. In what follows it will be more convenient to work instead in the Heisenberg picture where the state is stationary, $\hat{\rho}_{SR} (0)$, while the time dependence of an operator $\Op (t)$ is governed by the commutator  $\D \Op  (t)/ \D t =({\rm i}/\hbar)[\H,\Op (t)]$. Expectation values at time $t$ can then be obtained as 
\be
 \tr [\hat{\rho}_{SR} (t) \, \Op(0)] = \tr [\hat{\rho}_{SR} (0) \, \Op(t)] .
\ee
Using the standard commutation relations for the spin operators (and orbital angular momentum operators in general), $[\Scomp{n}{j}, \Scomp{m}{k}]= {\rm i}\hbar \, \delta_{mn} \sum_{k}\epsilon_{j k l} \Scomp{n}{l} $, and for the position/momentum operators, $[\Xwcomp{n}{j}, \Pwpcomp{m}{k}] ={\rm i}\hbar \, \delta_{nm} \, \delta_{jk} \, \delta(\omega-\omega')$, we obtain the following equations of motion for the spin operators $\Spin{n} (t)$
\begin{align} \label{eq:spin_motion}
	\frac{\D \Spin{n}}{\D t} = \Spin{n} \times 
	\left[\gamma \left(\Bext + \Benv{n} \right)+\sum_{m\neq n} \bar{\mathcal{J}}^{(nm)}  \Spin{m} \right], 
\end{align}
where $\bar{\mathcal{J}}^{(nm)}=(1/2)[\mathcal{J}^{(nm)}+(\mathcal{J}^{(mn)})^{\rm T}]$ is the symmetrized exchange tensor and $\Benv{n}= \int_0^{\infty}\d\omega\, \Cw{n}  \Xw{n}$ is a magnetic field operator generated by the reservoir oscillator positions $\Xw{n}$ at site $n$. 

In turn, the equations of motion for these operators are 
\be
	\frac{\D^{2} \Xw{n}}{\D t^{2}}+\omega^{2} \Xw{n} 
	=\gamma \, \CwT{n}  \Spin{n} ,\label{eq:oscillator_motion}
\ee
i.e. the reservoir oscillators are driven by the motion of the spins, with the (transposed) coupling tensors $\CwT{n}$ governing the degree of driving for each of the continuum of oscillators.
We assume retarded boundary conditions so that the reservoir responds only to the past behaviour of the spins. The retarded Green function, $\Green_{\omega}(t-t')=\Theta(t-t')\sin(\omega(t-t'))/\omega$  obeys $(\partial_{t}^{2}+\omega^{2})\Green_{\omega}=\delta(t-t')$,  and Eq.~\eqref{eq:oscillator_motion} can then be solved exactly by
\begin{multline}
	\Xw{n}(t)=\sqrt{\frac{\hbar}{2\omega}} 
	\left(\aw{n} \, {\rm e}^{-{\rm i}\omega t}+ \awd{n} \, {\rm e}^{+{\rm i}\omega t}\right)\\
	+\gamma\int_{\tin}^{\infty} \d t'\, \Green_{\omega}(t-t')\, \CwT{n}   \Spin{n} (t') . \label{eq:oscillator_solution}
\end{multline}
Here, $\aw{n}$ and $\awd{n}$ are (vectors of) bosonic ladder operators with their components obeying $[\awcomp{n}{j}, \awpdcomp{m}{k}]=\delta_{nm} \, \delta_{jk}\, \delta(\omega-\omega')$. Classically these correspond to the two integration constants for the differential equation \eqref{eq:oscillator_motion} which set the initial amplitude and velocity of the oscillator.  

Substituting the reservoir solutions \eqref{eq:oscillator_solution} into the equations of motion for the spins \eqref{eq:spin_motion}, we obtain the first  result: The Heisenberg--Langevin equation that governs  three-dimensional quantum spin dynamics under the influence of memory and coloured quantum noise is 
\begin{multline}
	\frac{\D \Spin{n}(t)}{\D t}= \Spin{n} (t) \times 
	\bigg[ \gamma \Bext +\sum_{m\neq n} \bar{\mathcal{J}}^{(nm)}  \Spin{m} (t) \\
	+ \gamma \, \bstoch{n}(t)  +\gamma^{2} \int_{\tin}^{t} \d t' \, \kernel{n}(t-t')  \Spin{n} (t') \bigg] .\label{eq:modified_llg}
\end{multline}
The term $\bstoch{n}(t)$ is a Hermitian magnetic noise operator for site $n$, 
\be 	\label{eq:mag_noise}
	\bstoch{n}(t)
	=\int_{0}^{\infty} \d\omega \, \sqrt{\hbar \over 2\omega} \,\, \Cw{n}  \bigg(\aw{n} \, e^{-{\rm i}\omega t}	+ h.c.  \bigg), \quad
\ee 
which plays the role of the stochastic noise first described by Brown~\cite{brown1963}. Here it {\it arises} from the spin's interaction with its reservoir. As we will see below, the bath noise can be coloured and contain quantum zero-point fluctuations.
In addition to the coloured noise $\bstoch{n}$, a kernel tensor $\kernel{n}(t-t')$ appears in Eq.~\eqref{eq:modified_llg}, which captures the damping of the spins. It arises from the coupling tensor $\Cwtens$ and is given by
\be	\label{eq:response_function}
\kernel{n}(\tau) = \Theta(\tau) \, \int_{0}^{\infty}\d \omega \, \frac{\Cw{n}  \CwT{n}}{\omega} \, \sin\left(\omega \tau \right),
\ee
where $\tau =t-t'$.
Here $\Theta$ is the Heaviside function which makes the spin's dynamics at time $t$, see \eqref{eq:modified_llg}, a function of the spin's state at previous times $\tin  \le t' < t$
\footnote{The Fourier transform $\tkernel{n}(\omega)$ of the kernel $\kernel{n}(\tau)$ automatically satisfies the Kramers--Kronig relations~\cite{volume5}, connecting the dissipative and reactive parts of the response kernel, as is required for any causal response.}.

\medskip

The three-dimensional spin dynamics equation \eqref{eq:modified_llg} describes the evolution of spin operators and explicitly includes memory of the past dynamics (non-Markovianity). This contrasts with previous derivations of quantum spin dynamics in the form of a master equation  \cite{nieves2014} which can be solved numerically. But to obtain the master equation a range of simplifying assumptions where made, including as weak spin-environment coupling, as well as the Markov and secular approximations. These approximations are quite strong and may not always be justified for a given spin system. 
%
We further remark that a different method of including (classical) coloured noise is based on the Miyazaki-Seki approach  \cite{Atxitia2009}. Similar to \eqref{eq:spin_motion}, this model assumes that the equation of motion of the spins is coupled to a stochastic equation of motion for the lattice enabling transfer of energy and angular momentum between the lattice and the spin systems. Different coupling potentials, such as harmonic and Morse potentials, have been considered and the spin-lattice coupling impact on the magnetisation has been characterised \cite{Strungaru2021}.

\medskip

The above spin+environment Hamiltonian and its resulting equations of motion share many similarities with the well-known Caldeira-Leggett model for harmonic quantum Brownian motion \cite{caldeira,philbinanders2016} and the spin-boson model  \cite{Thoss2001,ABV2007,Purkayastha2020}. A key difference is the three-dimensional nature of the reservoir interaction in \eqref{eq:Vint} which leads to the cross product in \eqref{eq:modified_llg}. The spin-boson model is recovered as a special case for spin $1/2$ operators and rank-1 coupling tensors, see \Appendix~\ref{sec:spin-boson}. 

\subsection{\bf Fluctuation-Dissipation Relation and $\Spinwo^2$}

The presence of the kernel in \eqref{eq:modified_llg} gives rise  to spin damping that can significantly differ from Gilbert damping, as explored further in section \ref{sec:couplings}. Previous generalisations to Gilbert damping have considered the spins' interactions with the lattice and electron motion~\cite{Bauer08,asmann2019, fahnle2019, Strungaru2021}, and included  inertial terms  \cite{ciornei2011} and further memory terms within a damping kernel~\cite{bose2011,schutte2014,thonig2015,bajpai2019}. First measurements have recently confirmed the presence of inertial corrections~\cite{li2015,Neeraj2021}. 
But these studies have not considered how the noise may have to change from the standard white magnetic noise that is  included in the effective magnetic field in the LLG equation~\eqref{eq:llg_equation}. 

A separate strand of reasoning has argued that to extend the LLG equation to the quantum setting requires that the stochastic noise included in the effective field $\boldsymbol{B}_{\rm eff}$ should have a quantum distribution rather than the classical Boltzmann one \cite{Oppeneer1998, woo2015,bergqvist2018,Barker2019,Barker2020}. These studies partially reproduce experimentally measured magnetization over temperature curves. But the noise considered vanishes at low temperatures, whereas the full Bose-Einstein distribution famously has non-zero quantum fluctuations even at T = 0 \Kelvin.
Other pioneering investigations have discussed the effect of coloured noise and non-Markovian effects on the ultrafast demagnetization rate~\cite{Atxitia2009}, while not requiring the fluctuation-dissipation theorem (FDT) to hold and not including quantum effects.

\medskip

The quantum spin dynamics equation \eqref{eq:modified_llg} brings these pieces, the damping kernel, quantum statistics and coloured noise, together in a quantum thermodynamically consistent framework. To see this we now prove that the reservoir's two signatures - the stochastic field $\bstoch{n}$ and memory kernel $\kernel{n}$ - always fulfil the (quantum) fluctuation-dissipation theorem, for any choice of coupling tensor $\Cw{n}$.

The reservoir power spectrum $\tilde{\PSD} (\omega)$ is defined as the (quantum symmetrised~\footnote{The autocorrelation function of two reservoir operators $A$ and $B^\dag$ in the thermal reservoir state $\hat{\rho}_R$ is defined as the expectation value of the Hermitian operator, ${\langle  \left\{A(t) , B^{\dag}(t-\tau) \right\} \rangle_{\beta}/2}$. In the classical case $A$ and $B^\dag$ commute at all times, removing the need for this distinction.}) expectation value of the autocorrelation function of the magnetic noise $	\bstoch{n}(t)$ in the thermal reservoir state $\hat{\rho}_R = e^{-\beta \HR}/\tr[e^{-\beta \HR}]$, and then taking the Fourier transform from the time to the frequency domain~\footnote{The Fourier transform is here defined as $\tilde{f} (\omega)	= \int_{-\infty}^{\infty} \d \tau \, e^{+{\rm i} \omega \tau} f(\tau)$, with the inverse $f(\tau) = \int_{-\infty}^{\infty} {\d \omega \over 2 \pi} \, e^{-{\rm i} \omega \tau}  \tilde{f} (\omega)$.}, i.e. 
\be \label{eq:PSDdef}	\tilde{\PSD}^{(nm)}_{jk} (\omega)
	= \int_{-\infty}^{\infty} \d \tau \, e^{{\rm i} \omega \tau} {\left\langle \left\{\bstochcomp{n}{j}(t), \bstochdcomp{m}{k}(t-\tau) \right\} \right\rangle_{\beta} \over 2},  \quad \quad
\ee
where $\{ . \, , . \}$ is the anti-commutator.
Inserting \eqref{eq:mag_noise}, one finds that the only non-trivial contributions to this expression come from the bosonic ladder operator expectation values $\langle \{ \awcomp{n}{l}, \awpdcomp{m}{l'} \} \rangle_{\beta}=\delta_{nm} \,\delta_{ll'} \, \delta(\omega-\omega') \coth(\beta \hbar\omega/2)$. Meanwhile by Eq.~\eqref{eq:response_function} the square of the coupling tensor, $\Cw{n}  \CwT{n}$, is related to the imaginary part of the Fourier transform of the damping kernel $\kernel{n}(\tau)$, i.e. 
\be
	\Cw{n}  \CwT{n} = {2 \omega \over \pi} \, \mbox{Im}[\tkernel{n} (\omega)] .\label{eq:coupling_kernel}
\ee 
Hence one obtains for {\it all} choices of the interaction tensor $\Cw{n}$ in Eq.~\eqref{eq:Vint} the quantum fluctuation-dissipation theorem (FDT)
\begin{subequations} \label{eq:FDT}
\be	\label{eq:FDTq}
	\tilde{\PSD}^{(nn)}_\quant (\omega)
	&=& {\hbar} \, \mbox{Im}[\tkernel{n} (\omega)] \,  \coth {\hbar\omega \over 2 k_B T}, \\
	\label{eq:FDTc} \tilde{\PSD}^{(nn)}_\class (\omega)	
	&=&   {2 k_B T \over  \omega}  \, \,  \mbox{Im}[\tkernel{n} (\omega )]  \, \mbox{ for } \,  k_BT \gg {\hbar\omega \over 2}, \quad \quad
\ee
\end{subequations}
where the last line is the well-known classical high-temperature (or low frequency) approximation, and the first line is the low-temperature limit where the Bose-Einstein distribution of the reservoir oscillator modes becomes important~\footnote{The power spectrum given in \eqref{eq:FDT} is the correct general version for any kernel $\kernel{n}(t)$, fulfilling the full quantum FDT~\cite{volume5}.  For a Gilbert damping kernel a power spectrum proportional to $\hbar\omega/(\exp{(\hbar\omega/k_B T)}-1)$ was given in~\cite{Oppeneer1998,woo2015,bergqvist2018,Barker2019}.  This is missing the quantum ground state contribution of $\hbar\omega/2$, which acts as stochastic noise on the spin system even at zero temperature.}. 

\smallskip

Finally, the spin dynamics equation \eqref{eq:modified_llg} leaves the square of the spin operators a constant of motion, since
\be
	\frac{\D \big|\Spin{n}\big|^2}{\D t}
	= {i \gamma \hbar \over 2} \sum_{l} [\Scomp{n}{l}, B_{{\rm eff},l}^{(n)}]
	= 0,\quad \quad
\ee
see \Appendix~\ref{app:constantS}. Here $\Beff{n} = \Bext + \Benv{n} +{1\over \gamma} \sum_{m\neq n} \bar{\mathcal{J}}^{(nm)}  \Spin{m}$ is an effective magnetic field at site $n$ see Eq.~\eqref{eq:spin_motion},  which commutes with all components of $\Spin{n}$.  This confirms that the spin length $|\Spinwo\big|$  is a constant in time, e.g. for a spin-1/2  $\Spinwo = {\hbar \over 2} \hat{\boldsymbol{\sigma}}$ with $\hat{\sigma}_j$ the Pauli matrices, one has a constant $\Spinwo^2={\hbar^2 \over 4}  \hat{\boldsymbol{\sigma}}^2 = {3 \hbar^2 \over 4} \, \mathbbm{1}_2$.

\section{Coupling functions} \label{sec:couplings}

The general quantum spin dynamics equation \eqref{eq:modified_llg} is specified completely by the coupling tensor $\Cwtens$ in \eqref{eq:Vint} that sets the interaction between spins and reservoirs.  
Here we will show that the standard LLG equation \eqref{eq:llg_equation} arises as a special case of \eqref{eq:modified_llg}, for a particular choice of coupling $\Cwtens$. 
We will further introduce a class of Lorentzian coupling functions which allow a systematic exploration of spin dynamics behaviours beyond the LLG equation. 

For simplicity from here on we will drop the lattice site superscripts and consider isotropic coupling tensors $\Cwtens =\Cwwo \, \mathbbm{1}_3$ with $\Cwwo$ a scalar function, and similarly for the corresponding kernels $\kernelwo (\tau)=K_{\omega}(\tau) \, \mathbbm{1}_3$ and power spectra $\tilde{\mathcal{P}}(\omega)=\tilde{P}(\omega)\mathbbm{1}_3$. These choices are appropriate for 3D materials in which all spins couple to the environment in the same manner in all spatial directions. For other materials, such as 2D layers, non-isotropic coupling tensors can be considered.

\subsection{Ohmic coupling and recovery of LLG equation} \label{sec:Ohmiccoupling}

In the open quantum systems literature, coupling functions that are linear in frequency are referred to as ``Ohmic'', while those proportional to higher and lower powers of $\omega$ are called super- and sub-Ohmic, respectively. For the magnetic system considered here, Ohmic coupling means 
\be
    \Cwwo^\Ohm=\, \sqrt{\frac{2\eta_{G}\cos(\omega\epsilon^{+})}{\pi}} \, \omega\label{eq:ohmic},
\ee
where $\eta_{G}$ is a positive constant with units of $\mbox{kg}/\mbox{A}^2 \mbox{m}^2  \sec$, as in Eq.~\eqref{eq:llg_equation}, and $\epsilon^{+}$ is an infinitesimal positive constant which we take to zero at the end of the calculation.  The corresponding kernel \eqref{eq:response_function} is close to instantaneous,
\begin{align}
    K^\Ohm(\tau)
    &= \frac{\eta_{G}}{\pi} \, \Theta(\tau)\int_{0}^{\infty}\d\omega\,\omega[\sin(\omega(\tau+\epsilon_{+}))\nonumber\\
    &\hspace{4cm}+\sin(\omega(\tau-\epsilon_{+}))]\nonumber\\
    &=-\eta_{G} \, \Theta(\tau) 
   \frac{\D}{\D\tau} \, \delta(\tau-\epsilon_{+}), \label{eq:ohmic_damping_kernel}
\end{align}
from which we can see that a positive value of $\epsilon^{+}$ ensures the response kernel $K^\Ohm$ is causal. For Ohmic coupling \eqref{eq:ohmic}, the damping term in \eqref{eq:modified_llg}  reduces to
\begin{multline}
    \gamma^2\Spinwo(t)\times\int_{\tin}^{t}\d t'\,K^\Ohm(t-t')\Spinwo(t')\\
    =-\gamma^2\eta_{G} \, \Spinwo(t)\times\frac{\D\Spinwo(t-\epsilon^{+})}{\D t}, \label{eq:LLGdamping}
\end{multline}
i.e. Ohmic coupling results in a damping term proportional to a first order time derivative of the system variable. In the magnetic case this is $\Spinwo$,  while for quantum Brownian motion  it would be the oscillator position \cite{caldeira}. Inserting \eqref{eq:LLGdamping} into the general quantum spin dynamics equation \eqref{eq:modified_llg}, we see that it recovers   (the quantum version of) the LLG equation \eqref{eq:llg_equation} with $\eta_G$ being identified as the damping parameter. This becomes the well--known classical equation in the limit of large spins.  Consequently we will also refer to the Ohmic coupling function (\ref{eq:ohmic}) as ``LLG coupling'' \footnote{Note that although \eqref{eq:LLGdamping} is not an explicitly  Hermitian operator, \Appendix~\ref{ap:hermiticity} shows that Eq.~\eqref{eq:modified_llg} is nevertheless equivalent to a Hermitian equation of motion.}.

\begin{figure*}[t]
\includegraphics[trim=0.45cm 0.5cm 0.35cm 0.35cm,clip, width=1.0\textwidth]{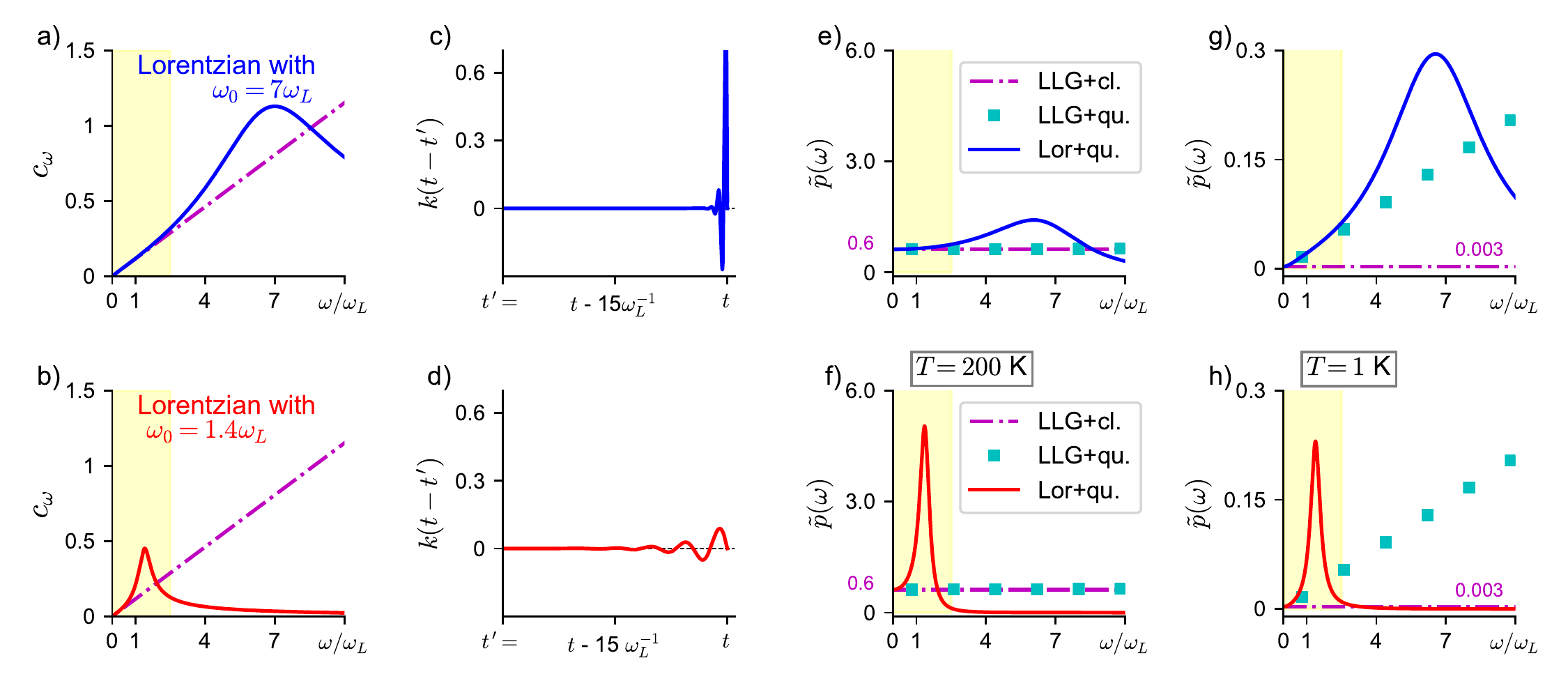}
\caption{\label{fig:PSD} \sf {\it Comparison of coupling functions, memory kernels and power spectra:} 
Top panels show  (a)  coupling function $c_{\omega}$, (c) time dependent damping kernel $k(t-t')$, and magnetic noise power spectrum $\tilde{p}(\omega)$ at (e) $T=200 \Kelvin$, and (g) $T=1 \Kelvin$, for Lorentzian coupling with parameter Set~1 \eqref{eq:sets1} (solid blue). 
Typical spin dynamics frequencies $\omega \in [0,2.5 \omega_L]$ are  highlighted (yellow shading) in the frequency domain.
Bottom panels shows the same quantities for  Lorentzian coupling with parameter Set~2 \eqref{eq:sets2}  (solid red). 
In (a--b) the Lorentzian coupling functions are compared to the LLG (Ohmic) approximation (magenta dash-dotted). 
In (e--h) the Lorentzian power spectrum is compared to the quantum LLG approximation (cyan squares), and its high temperature white noise limit (magenta dash-dotted). 
}
\end{figure*}

Given a coupling function it is straightforward to find the corresponding power spectrum \eqref{eq:FDTq} governing the time correlations of the quantum noise. For $\Cwwo^\Ohm$ the quantum power spectrum is 
\be
    \tilde{P}^\Ohm_\quant(\omega)=\, \eta_{G} \, \hbar\omega \, \coth\left(\frac{\hbar\omega}{2k_{B}T}\right), \label{eq:llg_psd}
\ee
where we have taken the limit $\epsilon^{+}\to0$.
The Ohmic quantum power spectrum \eqref{eq:llg_psd} is proportional to the damping parameter $\eta_{G}$, and tends to a linear (diverging) function of frequency in the low temperature limit $\hbar\omega\gg k_{B}T$, see Fig.~\ref{fig:PSD}g+h.  
However, power spectra diverging with frequency are unphysical. Coupling any system to  environmental modes must go to zero at large enough frequencies, such as the interaction of spins with lattice phonons or the effect of conduction electrons scattering off the spins. A common way of integrating this physical information into Ohmic coupling is to introduce a cut-off, i.e. an upper frequency to which the coupling grows linearly, and after which it is set to 0 or decays to 0 in either algebraic or exponential form \cite{Breuer-Petruccione}. A different approach was taken in  \cite{Oppeneer1998,Barker2019} for the modelling of spin dynamics,  where a semi--quantum Ohmic power spectrum similar to \eqref{eq:llg_psd} was chosen, but with the zero--point noise subtracted ensuring that at $T=0$K it vanishes for all frequencies.

In the high temperature limit $k_{B}T\gg\hbar\omega$, Eq.~\eqref{eq:llg_psd} reduces to the classical power spectrum
\be 
    \tilde{P}^\Ohm_\class(\omega) 
    = \, 2\eta_{G} \, k_{B}T,\label{eq:ohmic_psd_classical}
\ee
which is frequency independent (white) noise. Thus Ohmic coupling plus the high temperature approximation to the power spectrum recovers the LLG equation \eqref{eq:llg_equation} commonly used to simulate magnetic materials over a wide range of temperatures.

\medskip

As a first step to unravel how the dynamics predicted by the quantum spin equation \eqref{eq:modified_llg} can deviate from that predicted by the LLG equation \eqref{eq:llg_equation}, one can expand the spin vector (operator) at time $t'$ in Eq.~\eqref{eq:modified_llg} around the end point of integration to arbitrary order, 
\begin{multline}
    \gamma^2\Spinwo(t)\times\int_{\tin}^{t}\d t'\,K(t-t')\Spinwo(t')\\
    =\gamma^2\sum_{m=1}^{\infty}\kappa_m \, \Spinwo(t)\times\partial_{t}^{m}\Spinwo(t). \label{eq:damping_expansion}
\end{multline}
The $0$-th order coefficient, $\kappa_{0}$ does not appear in the expansion of the damping operator \eqref{eq:damping_expansion}, as $\Spinwo(t)\times\Spinwo(t)=0$. For the quantum operators $\Spinwo(t)$ this is ensured by the angular momentum commutation relations, which make this cross product anti--Hermitian. As outlined in \Appendix~\ref{ap:hermiticity}, when using any truncated form of expansion \eqref{eq:damping_expansion} one must employ an explicitly Hermitian form of Eq.~\eqref{eq:modified_llg}, and hence any anti-Hermitian contributions drop out.
The higher expansion coefficients $\kappa_{m>0}$ are proportional to the $m^{\rm th}$ one--sided moment of the memory kernel 
\be
    \kappa_m= \frac{(-1)^{m}}{m!}\int_{0}^{\infty}\d\tau  \, \tau^{m} \, K(\tau),
\ee
where we have assumed that the dynamics has been running for some time longer than the kernel decay time, for which one can replace the initial time  $\tin$ by $- \infty$.

For the Ohmic kernel only the first moment is non-zero and corresponds to the (negative) damping parameter, 
\be 
    \kappa_{1}^\Ohm 
    = \eta_{G} \int_{0}^{\infty}\d\tau \,   \tau   \, \frac{\D}{\D\tau} \, \delta(\tau-\epsilon_{+})  
    = - \eta_{G},
\ee 
while $\kappa_{m>1}^\Ohm=0$. This complete lack of higher moments, which would maintain a certain degree of memory in the dynamics, shows that Ohmic coupling dynamics can only be an approximation to any real dynamics.
For example within magnetism, the memory-free form of damping \eqref{eq:LLGdamping} is known as Gilbert damping~\cite{gilbert}, and is almost universally used to describe magnetization dynamics through the LLG equation \eqref{eq:llg_equation}. However, ``inertial'' corrections to such dynamics have been proposed \cite{ciornei2011} and their presence was recently  confirmed experimentally \cite{Neeraj2021}.

\subsection{Lorentzian coupling\label{sec:Lorentzian_coupling}}

Here we provide a tool to systematically study dynamics beyond the Ohmic case, allowing one to include memory and coloured noise effects in a manner consistent with the quantum fluctuation dissipation theorem \eqref{eq:FDTq}. We consider the class of Lorentzian coupling functions  
\be
    \Cwwo^\Lor =\sqrt{\frac{2A\Gamma}{\pi}\frac{\omega^2}{(\omega_0^2-\omega^2)^2 + \omega^2\Gamma^2}}, \label{eq:LorCw}
\ee
where $A$ is a coupling amplitude, with the following properties:  i) for small $\omega$, $\Cwwo^\Lor$ grows linearly with $\omega$ and can be approximated by an Ohmic coupling function, ii) at large $\omega$, $\Cwwo^\Lor$ smoothly decays to zero, and iii) at some intermediate frequency $\omega_0$, $\Cwwo^\Lor$ has a resonant peak with some width $\Gamma$. This peak characterises the confined range and relative strength of system-environment interaction with two parameters. Alternative ``peaks'' such as Gaussians or top hat functions could be considered, but here we chose the Lorentzian shape due to the fact that many expressions can be solved analytically and, as we will demonstrate in section \ref{sec:simulations}, Lorentzian couplings allow us to efficiently simulate non-Markovian dynamics.

We call the above functions ``Lorentzian coupling'' since the corresponding damping kernel in the frequency domain is the widely studied Lorentzian response 
\be
    K^\Lor(\omega)
    = \frac{A}{\omega_0^2-\omega^2-{\rm i}\omega \Gamma},\label{eq:Lorkernel}
\ee
where the imaginary part is obtained from \eqref{eq:coupling_kernel} and the real part is determined using the usual Kramers--Kronig relations.
In the time-domain the Lorentzian memory kernel is
\be \label{eq:Lorkern}
    K^\Lor(\tau) = \Theta(\tau) \, A \, e^{- {\Gamma \tau \over 2}} \, {\sin(\omega_1 \tau) \over  \omega_1},
\ee
where $\omega_1 = \sqrt{\omega_0^2 -{\Gamma^2 \over 4}}$ and $\Gamma/2$ can now be interpreted as the kernel decay rate.  For the coupling function \eqref{eq:LorCw}, the collective response of the environment is thus equivalent to a single harmonic oscillator  of resonant frequency $\omega_0$ and damping rate $\Gamma/2$~\cite{Correa2019}.
From the quantum FDT \eqref{eq:FDTq} it follows that the corresponding power spectrum  is 
\begin{equation}  \label{eq:Lorpow}
    \tilde{P}^\Lor_\quant (\omega) 
    = \frac{A\Gamma \hbar \omega}{(\omega_0^2-\omega^2)^2 + \omega^2\Gamma^2}\coth\left(\frac{\hbar\omega}{2k_{B}T}\right),
\end{equation}
which takes its largest values at frequencies close to $\omega_0$ and tends to zero as $\omega^{-3}$ at large $\omega$. This power spectrum  differs from classical Ohmic  noise \eqref{eq:ohmic_psd_classical} in two important respects.  Firstly, the quantum mechanical treatment means that the low temperature noise is not proportional to temperature, and does not vanish at zero temperature.  Secondly, even in the high temperature limit, the noise spectrum is frequency dependent (coloured), unlike the white noise of Eq.~\eqref{eq:ohmic_psd_classical}.  The presented theory thus captures both of these aspects in a consistent quantum thermodynamic framework.

\medskip

Unlike Ohmic coupling \eqref{eq:ohmic} which depends on a single parameter $\eta_{G}$, the Lorentzian coupling function \eqref{eq:LorCw}, and hence its kernel and spectrum, depends on three parameters. These allow a systematic study of different regimes of the environment response. Specifically, the memory time of the environment can be continuously varied by changing $\omega_0$ and $\Gamma$, which can lead to very different magnetic behaviour. Beyond the spin dynamics explored here, the proposed Lorentzian coupling may also be a useful tool for the characterisation of quantum Brownian motion of a variety of systems, including oscillators and free particles.

\subsection{Two coupling regimes}

To better understand the relation between the Ohmic and Lorentzian coupling functions, one may consider their kernel moments $\kappa_{m}$ in expansion \eqref{eq:damping_expansion}.
In contrast to the Ohmic case, for the Lorentzian kernel \eqref{eq:Lorkern} all $\kappa_{m}$ are non-zero and given by
\be
    \kappa_m^\Lor
    =\frac{(-1)^m \, A}{\omega_1 \, \omega_0^{2(m+1)}} \, \, {\rm Im}\left[\left(\frac{\Gamma}{2}+{\rm i}\omega_1\right)^{m+1}\right]. \label{eq:kappan_Lor}
\ee
The first relevant two moments are
\be   
    \kappa_{1}^\Lor 
    =-\frac{A\Gamma}{\omega_0^4}
    \quad  \mbox{and} \quad
    \kappa_{2}^\Lor 
    =\frac{A}{\omega_0^{6}}(\Gamma^2 - \omega_0^2),  \label{eq:kappa012_Lor}
\ee
and when comparing to the Ohmic case, one finds that the first Lorentzian moment can be identified with minus the damping parameter $\eta_G$, i.e. $\kappa^\Lor_{1}=-{A\Gamma \over \omega_0^4}=-\eta_{G}$.
For a material with a given damping parameter $\eta_G$ this fixes one of the Lorentzian parameters, i.e.
\be \label{eq:lorcoupling2}
    \Cwwo^\Lor =\sqrt{\frac{2 \eta_G  \, \omega^2}{\pi}\frac{ \omega_0^4}{(\omega_0^2-\omega^2)^2 + \omega^2\Gamma^2}}, 
\ee
which now only depends on the two parameters $\omega_0$ and $\Gamma$. For a specific material these may be approximately determined through information contained in the density of states of the environment to which the spins couple~\cite{Nemati2021}. 

Inserting expansion \eqref{eq:damping_expansion} with  Lorentzian moments \eqref{eq:kappan_Lor} in the quantum spin dynamics equation \eqref{eq:modified_llg} one can distinguish two different dynamical situations.

\smallskip

\noindent {\bf Ohmic regime:} When the resonant frequency $\omega_0$ and the damping rate $\Gamma$ of the reservoir coupling is much larger than the spin operators' typical frequency of motion, $\omega_0\gg\omega$, each successive term in expansion \eqref{eq:damping_expansion} is smaller by an extra factor of $(\omega/\omega_0)$.  In the limit of infinite $\omega_0$ but finite damping parameter $\eta_G$, the Lorentzian damping term in \eqref{eq:modified_llg} thus tends to the Ohmic one \eqref{eq:LLGdamping}, $-\gamma^2\eta_{G} \, \Spinwo(t)\times \partial_t \Spinwo(t)$.

\smallskip

\noindent {\bf Non-Ohmic regime:} When the environmental frequency $\omega_0$ is comparable to typical spin motion frequencies, $\omega_0 \approx \omega$ the Ohmic approximation to the Lorentzian kernel begins to fail. The first deviation in \eqref{eq:modified_llg} is a new term proportional to $\kappa_{2}^\Lor$. I.e. in addition to the Gilbert damping term, one adds the term  $\gamma^2 \, \kappa_2^\Lor \, \Spinwo(t)\times\partial_{t}^{2}\Spinwo(t)$ containing a second time derivative of the spin operator $\Spinwo$. By analogy with the classical equation of motion for a massive body, this is known as an ``inertial'' modification to the spin dynamics \cite{ciornei2011}. As the ratio $\kappa_{2}/\kappa_{1}$ has the dimensions of time, one may introduce an ``inertial timescale'' $\tau_{\rm in}$~\cite{ciornei2011}, which for the Lorentzian is
\be
    \tau_{\rm in}=\frac{\kappa_{2}^\Lor}{\kappa_{1}^\Lor}=\frac{\omega_0^2-\Gamma^2}{\omega_0^2\Gamma}.
\ee
A large inertial time $\tau_{\rm in}$ indicates the presence of non-Markovian dynamics, i.e. dynamics that has a certain degree of memory.
For a high quality factor resonance $\omega_0 \gg \Gamma$ the inertial time  is (half) the kernel's decay time, $\tau_d =2/\Gamma$.  For increasing resonance width $\Gamma$, the inertial time $\tau_{\rm in}$ decreases and memory effects become less important.
In magnetic systems this timescale determines the time over which nutation oscillations are observed in the precession of the spin.  Such inertial corrections to standard magnetism have very recently been observed for the first time in ultrafast experiments on thin films~\cite{Neeraj2021}. Curiously, the inertial timescale becomes negative when $\Gamma > \omega_0$, a fact that may be the subject of future investigation.

Similarly to the kernel expansion \eqref{eq:damping_expansion}, one can also expand the Lorentzian power spectrum in frequency, see \Appendix~\ref{app:powerspecexpansion}. Generally, only the odd moments $\kappa_{2m+1}$ appear in the power spectrum. This implies that if a  kernel only has non-trivial first ($\kappa_1$) and second  ($\kappa_2$) moments, while higher moments vanish ($\kappa_{m>2} =0$) then the power spectrum will still be given by the (quantum) Ohmic one~\eqref{eq:llg_psd}, with $\eta_G = -  \kappa_1$. When third or higher moments are  non-zero, then the power spectrum of the noise will deviate from the Ohmic case at all temperatures.


To summarise, here we have demonstrated that Lorentzian coupling functions, kernels and power spectra provide a systematic framework to explore system dynamics that arises from inertial terms and other memory  effects, while recovering the standard Ohmic limit whenever the Lorentzian resonance frequency $\omega_0$ is much larger than the typical system frequencies.


\subsection{\bf Unit-free variables and Lorentzian parameters} \label{subsec:Lorpara}

In section \ref{sec:simulations} we perform semi--classical simulations of the dynamics of Eq.~\eqref{eq:modified_llg}, for the Lorentzian kernel of Sec.~\ref{sec:Lorentzian_coupling}.  For this purpose we re--write expressions in the operator equation \eqref{eq:modified_llg} in terms of a unit-free set of quantities.  The time coordinate $t$ is replaced with the unit-free coordinate $\omega_L t$, where $\omega_{L}=|\gamma||\Bext|$ is the Larmor frequency.  In addition the spin operator $\Spinwo$ with largest eigenvalue $S_0$ is re--written in terms of a unit-free operator $\hat{\boldsymbol{s}}$ with largest eigenvalue $1$, $\Spinwo={\rm sign}(\gamma)\,S_0\,\hat{\boldsymbol{s}}$.  The sign of the gyromagnetic ratio is included in the definition of $\hat{\boldsymbol{s}}$ so that $\hat{\boldsymbol{s}}$ aligns with the magnetic field, whatever the sign of $\gamma$.  

From Eq.~ \eqref{eq:modified_llg} we can see that the damping kernel has dimensions of magnetic field squared divided by angular momentum, which leads us to identify a unit-free damping kernel $k(t-t')$ through $K(t-t')=|\Bext|^2 \, S_0^{-1} \, k(t-t')$.
Similarly, the unit-free coupling function $c_{\omega}$ is defined through $\Cwwo=|\Bext| \, S_0^{-1/2} c_{\omega}$ and the unit-free spectral functions $\tilde{p}$ through $\tilde{P}=\hbar |\Bext|^2 \, S_0^{-1} \, \omega_L^{-1}  \tilde{p}$.


Looking at the Lorentzian kernel $K(t-t')$ in \eqref{eq:Lorkernel}, the pulling of dimensions can be achieved by setting the kernel amplitude to  $A=|\Bext|^2 S_0^{-1}\alpha$ where  $\alpha$ now is a  frequency. For the simulations we choose the Lorentzian parameters $\omega_0, \Gamma$ and $\alpha$ to be independent of spin length $S_0$. Through \eqref{eq:Vint} this implies an interaction energy scaling of $\Vint \propto S_0 \, \sqrt{A} \propto \sqrt{S_0}$ which sets the scaling of the interaction versus self-energy to $\Vint/\HS \propto 1/\sqrt{S_0}$. Apart from it being implied by dimensional analysis, such scaling is heuristically plausible in many physical contexts. E.g. it is similar to the increasing ratio of the surface (where reservoir interaction occurs) to volume (self-energy) for decreasing system sizes. For microscopic systems for which $\Vint$ is no longer small in comparison to $\HS$ a thermodynamic treatment beyond the weak coupling limit \cite{Miller2018,Jarzynski2017,philbinanders2016}, a limit tacitly assumed in standard thermodynamics, may be required. 

Similarly for  Ohmic coupling leading to the LLG equation \eqref{eq:llg_equation}, the above scaling choice amounts to choosing $\eta_G \propto A \propto 1/S_0$ implying that $\eta = \eta_{G} \gamma^2 S_0 = \alpha \Gamma \omega_L^2/\omega_0^4$ is assumed to be independent of $S_0$.   In physical situations where this assumption is not justified, one may instead choose $\eta$ and $\alpha$ to depend on the spin length $S_0$.

\begin{figure*}[t]
\includegraphics[trim=0.2cm 0.2cm 0.2cm 0.2cm,clip, width=0.95\textwidth]{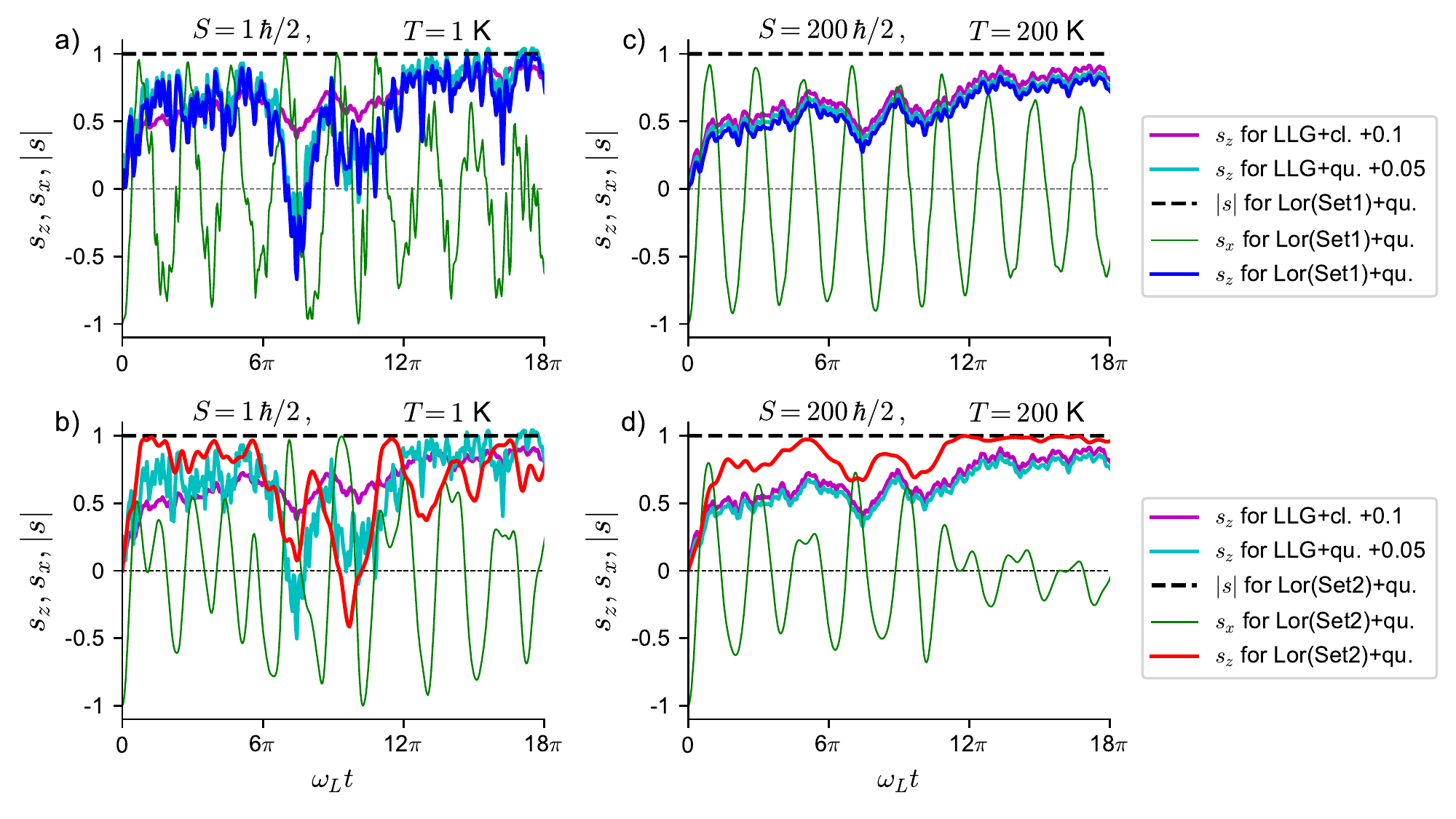}  
\caption{\label{fig:shorttimespindyn-4x4}   
\sf {\it Sample of stochastic short-time spin dynamics for different couplings and noises.} 
Stochastic short-time dynamics of $s_z$ (blue in top row \& red in bottom row), $s_x$ (green) and $|\boldsymbol{s}|$ (black dashed) according to Eq.~\eqref{eq:modified_llg} with Lorentzian coupling function $\Cwwo^\Lor$,  for a classical spin initially in state $\boldsymbol{s} = (-1,0,0)$. 
All traces in the four panels are generated from the same sample of Gaussian noise, enabling a direct comparison.
Shown are the dynamics for Set~1 (top row) and Set~2 (bottom row), and two spin+temperature pairs:  $S_0 = 1 \hbar/2$ and $T=1 \Kelvin$ (left column), and $S_0 = 200 \hbar/2$ and $T=200 \Kelvin$ (right column). 
Also shown in all four panels are the $s_z$-dynamics according to the LLG equation Eq.~\eqref{eq:llg_equation} with damping parameter $\eta  \approx 0.02$ for two types of noise: the classical flat white noise power spectrum Eq.~\eqref{eq:ohmic_psd_classical} (magenta) and the quantum noise power spectrum Eq.~\eqref{eq:llg_psd} (cyan). 
All cyan and magenta plots are off-set by +0.05 and +0.1, respectively, to avoid overlapping. 
The external magnetic field is set to $\Bext=(0,0,10\Tesla)$ setting the  timescale to $\omega^{-1}_L \approx  0.57  \cdot 10^{-12} \sec$, and the simulation time interval is $\D t= 0.15 \omega_L^{-1}$. 
}
\end{figure*}

For a spin in an external field $\Bext$ the typical frequency of the dynamics is set by the Larmor frequency, $\omega_L$. In the simulations discussed in section \ref{sec:simulations} we will use the following two sets of Lorentzian parameters, all expressed in terms of $\omega_L$, 
\begin{subequations}
\label{eq:sets}
\be
    \mbox{Set~1): }
    &\begin{matrix}
     \omega_0 = 7.0\,\omega_L   
     & \alpha = 10.0\,\omega_L 
     & \Gamma = 5.0\,\omega_L\\
     \eta= 0.02   
     & \tau_{\rm in} = 0.1\, \omega_L^{-1} 
     & \tau_d = 0.4 \,\omega_L^{-1}
    \end{matrix}  \quad \quad \,\, \label{eq:sets1}\\[5pt]
    \mbox{Set~2): }
    &\begin{matrix}
    \omega_0 = 1.4\,\omega_L 
    & \alpha = 0.16\,\omega_L  
    & \Gamma=0.5\,\omega_L\\ 
    \eta= 0.02 
    & \tau_{\rm in}=1.7 \, \omega_L^{-1} 
    &  \tau_d = 4 \,\omega_L^{-1} 
    \end{matrix}. \quad \quad \,\, \label{eq:sets2}
\ee
\end{subequations}
In the second rows we have also listed the equivalent unit-free Gilbert damping $\eta$, the inertial timescale $\tau_{\rm in}$, and the memory kernel decay time $\tau_{\rm d}$  for the Lorentzian kernel. 
Figs.~\ref{fig:PSD}-\ref{fig:MoverT}, show plots obtained with Lorentzian coupling functions \eqref{eq:LorCw} with parameter Sets 1 and 2, which are shown in blue and red, respectively.  The Ohmic LLG approximation is shown in magenta when the classical reservoir \eqref{eq:ohmic_psd_classical} is considered, and in cyan when the quantum reservoir \eqref{eq:llg_psd} is considered. The external field is set to $\Bext = 10\Tesla \, \boldsymbol{e}_z$ with $\boldsymbol{e}_z$ the unit-vector in $z$-direction throughout.


Parameter Set 1 has been chosen to have a resonant frequency $\omega_0$ much larger than the characteristic spin precession frequency $\omega_L$ (Ohmic regime). Consequently we can truncate the series given in Eq.~\eqref{eq:damping_expansion} to leading order and recover the Ohmic form \eqref{eq:LLGdamping} typically considered in magnetism theory.  The validity of this approximation is demonstrated in the top row of Fig.~\ref{fig:PSD}.  Fig.~\ref{fig:PSD}a) shows that the Lorentzian coupling function $c^\Lor_{\omega}$ is well approximated by LLG (Ohmic) coupling for the relevant frequency range, while Fig.~\ref{fig:PSD}c) shows that the kernel is approximately instantaneous on the timescale $\omega_L^{-1}$, in line with Ohmic damping for which $\kappa_{m>1}^\Ohm=0$.  Fig.~\ref{fig:PSD}e) shows that at high temperature ($T=200\Kelvin$) and for relevant frequencies $\omega \approx \omega_L$, the power spectrum $\tilde{p}^\Lor$ is well approximated by the quantum Ohmic (LLG) power spectrum \eqref{eq:llg_psd}, and its classical limit \eqref{eq:ohmic_psd_classical}.  Fig.~\ref{fig:PSD}g) shows that at lower temperatures ($T=1\Kelvin$) quantum noise becomes important.  Here the Ohmic approximation \eqref{eq:llg_psd} remains valid, while its classical limit \eqref{eq:ohmic_psd_classical} is invalid.  

Parameter Set~2 is chosen such that it has a resonant frequency $\omega_0$ comparable to the precession frequency $\omega_L$ (non-Ohmic regime). Here it is inaccurate to truncate the series \eqref{eq:damping_expansion} and the damping will be fundamentally non--Ohmic.  To directly compare with  Ohmic dynamics generated by Lorentzian coupling with  Set~1, both parameter sets have been chosen to correspond to the same unit-free Gilbert damping parameter, $\eta=0.02$. The failure of the Ohmic approximation is demonstrated in the bottom row of Fig.~\ref{fig:PSD}.  Fig.~\ref{fig:PSD}b) shows that the linear approximation to the coupling function fails in the relevant frequency range.  For this set of parameters the damping kernel now exhibits significant memory and Fig.~\ref{fig:PSD}d) shows that the response persists over a timescale of several $\omega_L^{-1}$. This memory kernel implies, through the FDT \eqref{eq:FDT}, a {\it coloured} quantum noise power spectrum $\pow^\Lor_\quant$. As shown in Fig.~\ref{fig:PSD}h), this coloured Lorentzian $\pow^\Lor_\quant$ (red) differs from the LLG counterpart, $\pow^\Ohm_\quant$ (cyan). Furthermore, Fig.~\ref{fig:PSD}f) shows that in the high temperatures $\pow^\Lor_\quant$ (red) also differs very significantly from the LLG power spectrum $\pow^\Ohm_\class$ (magenta). 
The presence of both memory and coloured quantum noise for the Lorentzian with parameter Set~2 are both signatures of a thermostat that substantially deviates from the classical Ohmic assumptions and, as we will see in the next section, leads to markedly different short time dynamics and steady state of $s_z$.

%

\section{\bf Semi-classical spin dynamics simulations} \label{sec:simulations}

The general spin dynamics equation \eqref{eq:modified_llg} is an operator equation for  quantum spins in a lattice, each interacting with neighbouring spins and with a bosonic reservoir. Solving the quantum dynamics using, for example, Lorentzian coupling, kernel and spectrum, is rather difficult without approximations, even numerically, and such exploration is left for future work.

To make progress here, we will numerically solve the full non-Markovian for a semi-classical version of Eq.~\eqref{eq:modified_llg}, while including coloured quantum noise and memory effects arising from the coupling to the environment.
%
%
It replaces the quantum spin operator $\Spinwo$ with a classical spin vector $\clspin$, and the quantum stochastic noise field vector $\bstochwo$ with a stochastic classical field vector $ \boldsymbol{b}$ with statistics that obey the quantum fluctuation--dissipation theorem \eqref{eq:FDTq}. 
This semi--classical approach is currently used in the theory of molecular and ionic dynamics~\cite{lu2012,lu2019}, and was perhaps first applied by Koch~\cite{koch1980} to include the effects of quantum fluctuations in Josephson junctions.  It has been justified through an expansion of a forward--backward path integral~\cite{schmid1982,kleinert1995} (note the remark of Caldeira and Leggett on pg. 589 of~\cite{caldeira}), and is valid when the potential energy can be expanded in the path integral to first order in the deviations from the average path. The validity of applying this approach to the decay of metastable states was investigated in detail in~\cite{eckern1990}.  

Here we simulate a single spin allowing us to illustrate the impact of memory effects and the reservoir's quantum statistics on the spin dynamics and steady state. The simulation details presented below can readily be extended to multiple interacting spins and could be integrated in sophisticated atomistic spin dynamics simulations such as those used in \cite{evans2014,Barker2019}.

\subsection{\bf How to simulate coloured noise and memory kernel} 

Here we detail how to efficiently simulate non-Markovian dynamics that arises as a result of Lorentzian coupling \eqref{eq:LorCw} for the example of spins vectors.
Numerical implementation of \eqref{eq:modified_llg} requires both - the integration of the kernel with the spin state of previous time steps and the inclusion of coloured noise as follows.

\smallskip

Dropping the spin index and for simplicity assuming any isotropic kernel $\kernelwo(\tau)=\boldsymbol{1}_{3} \, K(\tau)$, the three vector components $b_j (t)$ for $j=1,2,3$ of the magnetic noise \eqref{eq:mag_noise} are implemented as \cite{SchmidtMeistrenko2015}
\be \label{eq:noisegeneration}
    b_j (t) =\int_{-\infty}^{\infty} \d t' \, F(t-t') \, \xi_j(t'),
\ee 
where $\xi_j (t')$ is standard white Gaussian noise for the $j$-th component, which is delta correlated $\langle \xi_j (t) \, \xi_k(t')\rangle= \delta_{jk} \, \delta(t-t')$. 
The ``coloured noise" comes from choosing $F(t-t')$ as the Fourier transform of the square root of the power spectrum associated with the kernel through \eqref{eq:FDT}, i.e. 
\be \label{eq:noisegeneration2}
    F(t-t') = \int_{-\infty}^{\infty} {\, \d \omega \over 2\pi}  e^{- i \omega (t-t')}  \sqrt{ \tilde{P}  (\omega)},
\ee 
which can be implemented using a fast Fourier transform.
To simulate the effect of a Lorentzian damping kernel \eqref{eq:Lorkernel} we numerically integrate~\cite{scipy} the following set of first order coupled differential equations for the spin vector $\clspin$ and two dummy vectors $\boldsymbol{V}$ and $\boldsymbol{W}$: 
\be 
    \frac{\D \clspin (t)}{\D t} 
    &=&\gamma \clspin(t) \times (\Bext + \boldsymbol{b}(t) + \boldsymbol{V}(t)), \nonumber \\ 
    \frac{\D \boldsymbol{V}(t)}{\D t}
    &=&\boldsymbol{W}(t), \\ 
    \frac{\D \boldsymbol{W}(t)}{\D t}
    &=& - \omega_0^{2} \boldsymbol{V}(t) - \Gamma \boldsymbol{W}(t) + A  \, \gamma \clspin(t). \nonumber
\ee
The integrated values of the dummy vectors and the spin are separated by the time step $\D t$.  Solving these equations is equivalent to solving the integro--differential equation (\ref{eq:modified_llg}) for a Lorentzian kernel, see \Appendix~\ref{sec:kernel}, but is numerically more straightforward to implement.

\subsection{\bf Single trajectories for different couplings and noises.}

We wish to illustrate on a single trajectory level, the differences between the dynamics predicted by \eqref{eq:modified_llg} with either an approximately Ohmic (Set 1) or non-Ohmic (Set 2) Lorentzian coupling function, as well as the dynamics predicted by the standard LLG equation. At first, because the dynamics is intrinsically stochastic, trajectories will naturally differ and cannot readily be compared. However, looking at the noise generation in Eqs.~\eqref{eq:noisegeneration} and \eqref{eq:noisegeneration2}, one can see that the same  white noise $\xi_j(t)$ for $j=1,2,3$ may be used as a seed to create comparable ``stochastic" noise for different power spectra $\tilde{P}(\omega)$. 

Fig.~\ref{fig:shorttimespindyn-4x4} shows the stochastic short time dynamics of a single classical spin for two pairs of spin length and temperature, $S_0=1\hbar/2$ at $T=1 \Kelvin$ (left panel) for a single electron, and $S_0=200 \hbar/2$ at $T=200 \Kelvin$ (right panel) for a mesoscopic cluster of spins with a combined larger effective spin.

The dynamics is obtained according to Eq.~\eqref{eq:modified_llg} for Lorentzian coupling \eqref{eq:LorCw} with $S_0$-scaling $A=|\Bext|^2 S_0^{-1}\alpha$, for parameter sets Set~1 (top panel, blue) and Set~2 (bottom panel, red), and with the quantum coloured noise given by \eqref{eq:Lorpow}. 
For comparison we also show the short time dynamics according to the LLG equation \eqref{eq:llg_equation} with $S_0$-scaling $\eta_{G}=A\Gamma/\omega_0^4=|\Bext|^2 S_{0}^{-1}\omega_L^{-2} \eta$ with the Gilbert damping parameter $\eta =0.02$ common to both Lorentzian parameter sets, see \eqref{eq:sets}. That implies that the top and bottom LLG plots are identical. For the standard LLG equation two types of noise are considered - high-temperature classical noise (magenta) see Eq.~\eqref{eq:ohmic_psd_classical}, and quantum noise (cyan) see Eq.~\eqref{eq:llg_psd}. 
Since the same white noise time series is used as a seed for producing the stochastic noise for all traces, we can compare them directly. We will here focus on $s_z$, the  component of $\boldsymbol{s}=\mbox{sign}(\gamma) \clspin/S_0$ aligned with the external field $\Bext$. 

\begin{figure*}[t]
\includegraphics[trim=0.25cm 0.45cm 0.49cm 0.2cm,clip, width=0.95\textwidth]{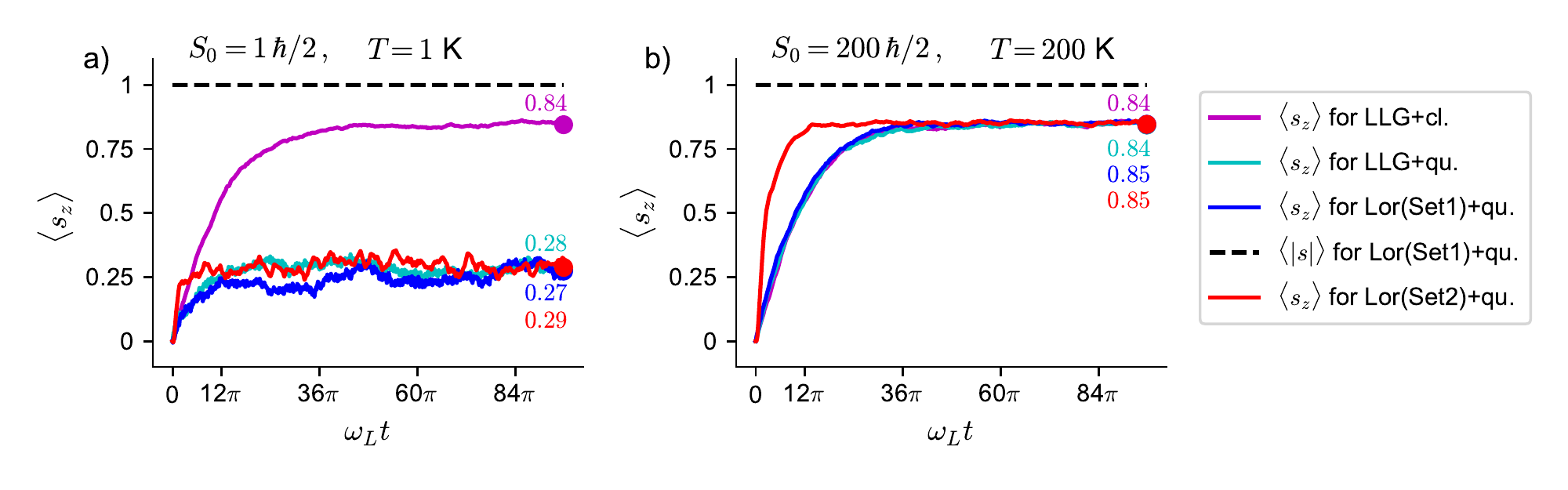} 
\caption{\label{fig:ensembleavg}  
\sf {\it Ensemble averaged spin relaxation dynamics.} 
Averaged dynamics $\langle s_z \rangle$, averaged over 500 stochastic traces up to time $t_{\max} = 2\pi \times 48 \omega_L^{-1}$.  Shown are the dynamics according to Eq.~\eqref{eq:modified_llg} with Lorentzian coupling functions for parameter Set~1 (blue) and Set~2 (red).
Also shown is the ensemble averaged dynamics according to the LLG equation with damping parameter $\eta  \approx 0.02$ for two types of noise:  classical white noise (magenta) and full quantum noise (cyan). 
The two panels show two spin+temperature pairs:  $S_0 = 1 \hbar/2$ and $T=1 \Kelvin$ (left), and $S_0 = 200 \hbar/2$ and $T=200 \Kelvin$ (right). Note that blue, cyan and magenta curves lie on top of each other in b). 
%
%
As discussed in section \ref{subsec:Lorpara}, the plots assume that both $A$ and $\eta_G$ scale as $\propto S_0^{-1}$ with spin size $S_0$, making $\alpha$ and $\eta$ the same for the two spin sizes.
The external magnetic field is set to $\Bext=(0,0,10\Tesla)$ and the simulation time interval is $\D t = 0.15 \omega_L^{-1}$.
}
\end{figure*}

Three features stand out in Fig.~\ref{fig:shorttimespindyn-4x4}: 
i) as expected from section \ref{sec:couplings}, the dynamics generated with Eq.~\eqref{eq:modified_llg} with Lorentzian Set~1  (top, blue) closely matches the dynamics obtained with the LLG equation \eqref{eq:llg_equation} with quantum noise (cyan) for both spin-temperature pairs, 
ii) the quantum statistics of the reservoir (cyan) at low temperatures (left) introduces differences to the LLG dynamics compared to the LLG dynamics obtained with classical noise (magenta), and  
iii) memory effects that are present for Lorentzian Set~2  (bottom, red) result in  significantly different dynamics from that arising with the memory-free Lorentzian Set~1  (top, blue).

We remark that due to the spin/temperature ratio being the same for the two spin-temperature pairs, the LLG equation with classical noise (magenta) integrates to exactly the same dynamics in left and right panel, see \Appendix~\ref{app:scaling}. This scaling relation ceases to be true for the LLG equation with quantum noise (cyan). Another difference to note is that in Fig.~\ref{fig:shorttimespindyn-4x4}a--d) the dynamics for Lorentzian parameter Set~1 (blue) varies more rapidly in time than for Lorentzian parameter Set~2 (red). This is due to the high frequency content of Set~1's power spectrum, see Fig.~\ref{fig:PSD}e+g).  

Finally, the spin component $s_x$ (green)  and the spin-vector length  $|\boldsymbol{s}|$  (black) are shown for the Lorentzian coupling with Set~1 and Set~2 in  the top and bottom panels of Fig.~\ref{fig:shorttimespindyn-4x4}, respectively. The plots of $|\boldsymbol{s}|$ show that the numerical integration of  Eq.~\eqref{eq:modified_llg} indeed leads to a constant spin-vector length $|\boldsymbol{s}|=1$,  i.e. no renormalisation is required.

\subsection{\bf Ensemble-averaged $\langle s_z \rangle$ trajectories.}
%

Fig.~\ref{fig:ensembleavg} shows the ensemble averaged  $\langle s_z\rangle$ over time, averaged over 500 stochastic trajectories. We now highlight two important features in Fig~\ref{fig:ensembleavg}. 
Firstly, at low temperatures (left)  the quantum statistics of the reservoir (blue, red, cyan) results in a much depleted value of $\langle s_z \rangle$, roughly at around $0.28$, in comparison to that obtained with the LLG equation with classical noise (magenta), ca $0.85$. This indicates that for this particular choice of spin length and temperature the quantum character of the reservoir strongly affects the value of $\langle s_z \rangle$, as further discussed below, and the classical high-temperature limit taken in \eqref{eq:FDTc} would not be appropriate. For the high temperature $T = 200 \Kelvin$ + larger spin pair $S_0 =200\hbar/2$ (right), the difference between classical and quantum statistics of the reservoir can be neglected and $\langle s_z \rangle$ settles at $0.85$ independent of whether the spin dynamics integration was done for Eq.~\eqref{eq:modified_llg} with either Set~1 (blue) or Set~2 (red), or  for Eq.~\eqref{eq:llg_equation} with either classical (magenta) or quantum noise (cyan).

Secondly, for the large spin-temperature pair (right), there is clear evidence of a much quicker relaxation to steady state (by a factor of a third) for Lorentzian Set~2 (red) compared to the other plots (blue, cyan, magenta). This is a non-Markovian effect that arises because the memory kernel for Set~2 has an appreciable memory over  time, see Fig.~\ref{fig:PSD}d), while the other memory kernels are (close to) instantaneous. This quicker equilibration occurs because the non-Markovian kernel leads to a smoother dynamics which in turn is more quickly sampled by the dynamical system. 

\begin{figure*}[t]
\includegraphics[trim=0.8cm 0.8cm 0.42cm 0.38cm,clip, width=0.99\textwidth]{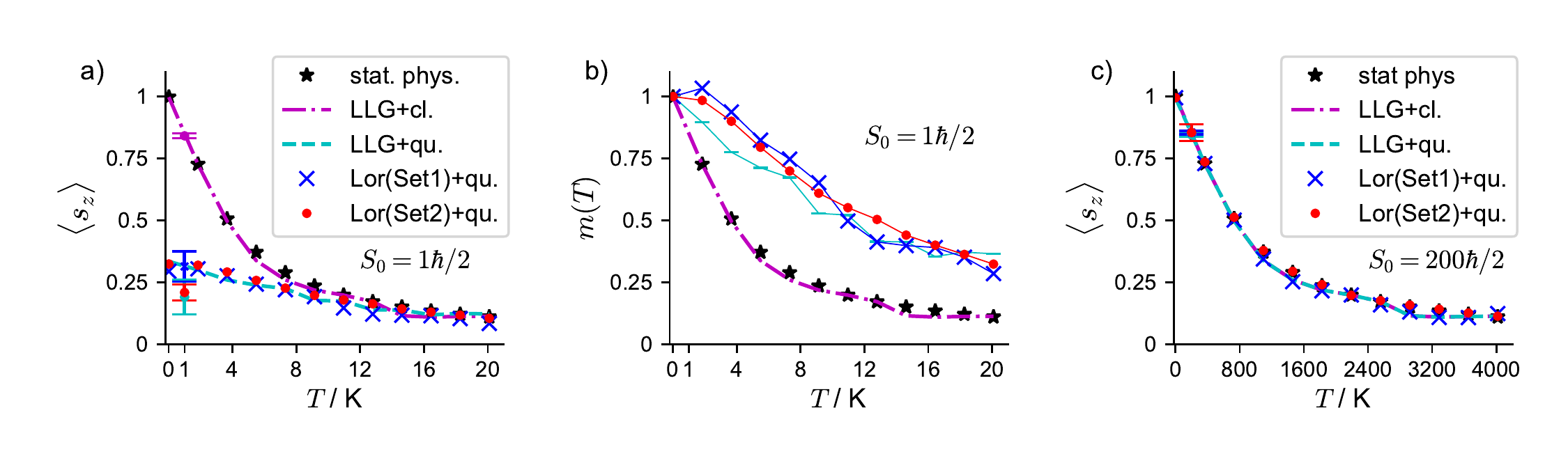}
\caption{ \label{fig:MoverT} 
\sf {\it Steady state $\langle s_z \rangle$ of a spin interacting with a (quantum) reservoir at temperature $T$.} 
For a spin in an external magnetic field  $\Bext=(0,0,10\Tesla)$ and interacting with a thermal reservoir, the time averaged value of $\langle s_z \rangle$ is obtained by integrating Eq.~\eqref{eq:modified_llg} for Lorentzian parameter Set~1 (blue crosses) and Set~2 (red dots), as well as by integrating the LLG equation with damping parameter $\eta = 0.02$ for two types of noise:  classical white noise (magenta dash-dotted) and full quantum noise (cyan dashed). 
The spin lengths are $S_0 = 1 \hbar/2$ (panel a and b), and $S_0 = 200 \hbar/2$ (panel c). 
For the small spin in panel a), the three curves with quantum noise (cyan, blue, red) start at $\langle s_z \rangle$-values in the range  0.2-0.4 at $T=0 \Kelvin$.
For both $S_0$ values, the steady state for the LLG equation with classical noise (magneta) coincide well with $\langle s_z \rangle_\stph$  (black), derived from classical statistical mechanics for a thermal distribution, see  \Appendix~\ref{app:clMT}.
Note that all curves lie on top of each other in panel c). 
Panel b) shows the same plot as panel a) - but with the $y$-axis ``rescaled'' as $m(T) =\langle s_z (T)\rangle/\langle s_z (0)\rangle$ so that all plots start at 1 at $T=0 \Kelvin$.
While the magenta curve remains the same as in a), the rescaled blue, red and cyan curves now show a flattened decay behaviour that somewhat resembles the corrected magnetization curves for real materials analysed in \cite{Evans15}.
Error bars for the four simulations (blue, red, cyan, magenta) are indicated at a) $T=1 \Kelvin$ and c) $T=200 \Kelvin$. 
The simulation time interval is $\D t = 0.15 \omega_L^{-1}$.
}
\end{figure*}

\subsection{\bf Steady state $\langle s_z \rangle$ as a  function of temperature.}


Fig.~\ref{fig:MoverT} shows the average steady state spin value $\langle s_z \rangle$ as a function of temperature $T$, found by  time-averaging a single trajectory $s_z$ over late times, from $0.75 \, t_{\rm max}$ to $t_{\rm max}=2 \pi \times 7200  \omega_L^{-1}$. 
There are two key observations to make in Fig.~\ref{fig:MoverT}. Firstly, for both the small spin (left) and the large spin (right) the steady state $\langle s_z \rangle$ obtained with the LLG equation and classical noise \eqref{eq:ohmic_psd_classical} matches the standard statistical physics prediction $\langle s_z \rangle_\stph =  \coth  \left({S_0  \omega_L  \over  k_B T} \right)  \, - {k_B T \over S_0 \omega_L}$, see \Appendix~\ref{app:clMT}. 

Secondly, for simulations that include the full quantum noise (cyan, blue, red) in the dynamics of the small spin at low temperatures (left), we observe reduced $\langle s_z\rangle$ values in  the range 0.2-0.4 at  $T=0 \Kelvin$, i.e., well below the classical value of 1. This arises because the power spectrum  $\tilde{P}^\Lor_\quant$, given through the quantum FDT \eqref{eq:FDTq}, includes quantum fluctuations which remain even for $T\to 0 \Kelvin$. 
The steady state curves with quantum noise also show a characteristic ``flattening'' compared to the steep decay of the steady state with temperature for classical noise (magenta).
Qualitatively speaking, this quantum zero point noise, when compared to classical noise, is as if thermal noise is ``on'' even at $0 \Kelvin$. I.e. taking the Larmor frequency as the relevant frequency, and setting $\tilde{P}^\Ohm_\quant (0 \Kelvin) = \tilde{P}^\Ohm_\class (T_\class)$ for the Ohmic coupling, for example, defines a classical temperature of $T_\class  = 6.7 \Kelvin$  ``equivalent'' to the quantum zero-temperature case.  
For $S_0=1\hbar/2$ the classical statistical physics steady state value at $T_\class$ is $\langle s_z \rangle_\stph \approx 0.3$. This indeed is of comparable size to the $\langle s_z\rangle$ values obtained with quantum noise at $T = 0 \Kelvin$. The corresponding steady state value for the large spin ($S_0 = 200 \hbar/2$) is $\langle s_z \rangle_\stph \approx 0.995 \approx 1$, see Fig.~\ref{fig:MoverT}c). 


Generally, for the  classical temperature $T_\class = {\hbar \omega_L \over 2 k_B}$ which is ``equivalent'' to the quantum zero-point noise, one obtains $\langle s_z \rangle_\stph = \coth  \left({2 S_0  \over \hbar} \right)  - {\hbar \over 2 S_0}$, which only depends on the spin length $S_0$ while being independent of the field strength $|\Bext|$. With increasing $S_0$ this function rises very sharply  from $\approx 0.3$ to $1$. For example for spin length $S_0 = 5 \hbar/2$, the quantum zero temperature value is $\langle s_z \rangle_\stph \approx 0.8$ and its decay with increasing temperate is shown in Fig.~\ref{fig:MoverT-appendix}b) in \Appendix~\ref{app:M(T)v2plot}. 

The middle panel, Fig.~\ref{fig:MoverT}b), gives an alternative illustration of the steady state value for the small spin as a function of environment temperature $T$. It shows the same plot as panel a),  but with the $y$-axis rescaled as $m(T) =\langle s_z (T)\rangle/\langle s_z (0)\rangle$. All plots now start at 1 at $T=0 \Kelvin$, independent of whether the dynamics was integrated with quantum or classical noise.  
Interestingly, the overarching behaviour of the resulting curves bears some resemblance with heuristically rescaled magnetization curves that match experimental data \cite{Evans15,Kuzmin}. Running high-end atomistic simulations of Eq.~\eqref{eq:modified_llg}, instead of \eqref{eq:llg_equation}, for  multiple interacting classical spins would answer if the quantum power spectrum's impact on their low temperature magnetisation behaviour as well as their Curie temperature is the reason for the apparent rescaling. 

We note that larger numerical uncertainties arise for the quantum noise because an additional scale is present in comparison to classical noise, see \Appendix~\ref{app:scaling}. Error bars obtained from an ensemble of simulations are indicated at one temperature value in a) and c) for all four $\langle s_z \rangle$ curves in Fig.~\ref{fig:MoverT}.

\section{\bf Conclusions and open questions} \label{sec:conclusion}

We have derived a general quantum spin dynamics equation, Eq.~\eqref{eq:modified_llg}, capable of describing three-dimensional precession and damping. The terms arising from the reservoir interaction are treated in a quantum thermodynamically consistent manner, by tracing the origin of both the memory kernel, $\kernelwo(\tau)$, and the stochastic noise, $\bstochwo(t)$, to a single coupling function, $\Cwwo$. 
Secondly, Lorentzian coupling functions were proposed and shown to provide a systematic means to investigate different dynamical regimes - from  Ohmic  to non-Ohmic dynamics which is subject to memory and coloured noise. We showed that {\it only} in the Ohmic regime, the standard LLG equation with Gilbert damping, widely used in magnetism, is recovered. 
Finally, we provided details of how to include  Lorentzian memory and coloured noise in numerical simulations of open system dynamics. For the example of a single spin  vector, we illustrated that a non-Ohmic Lorentzian kernel leads to a faster equilibration time of $\langle  s_z \rangle$ in comparison to the Ohmic (LLG) regime. We also discussed the steady state differences that arise when the full quantum thermostat with quantum zero-point noise is employed, in contrast to  classical white noise. 

The above three ingredients provide a complete framework for the simulation of damped three-dimensional precession including memory and coloured noise. It can readily be adapted in atomistic spin dynamics simulations \cite{evans2014,Barker2019} that solve the dynamics of millions of interacting spins.



The theory presented here will be a useful tool for investigating non--Markovian behaviour, opening up a number of avenues for future research at the intersection of quantum thermodynamics, magnetic materials and beyond.
For example, it is an open question to clarify the connection between the three-dimensional precession described by the spin equation~\eqref{eq:modified_llg} and  rotational Brownian motion. The orientation of a non-symmetric rotating body behaves analogously to the three-dimensional spin vector, and the motion and viscosity of a gas surrounding a rotating body simultaneously act on its motion while obeying the FDT as discussed in recent work~\cite{Kuhn2017,Stickler2018}.

Within magnetism, for particular materials of interest, detailed models of the coupling functions $\Cwwo$ can be developed that are based on an understanding of the interactions between spins, phonons, and electrons in the material \cite{Nemati2021}.  Coupling to optical modes may further be included to describe, for example, whispering gallery photon-magnon coupling which leads to an effective Gilbert damping term that can take either sign~\cite{Kusminskiy2016}.
A direct experimental characterisation of a material's damping kernel $\kernelwo$ that determines memory and noise in \eqref{eq:modified_llg} may be attempted, for example with high field experiments such as those recently reported in~\cite{Neeraj2021}. 
Of particular interest are dynamical features beyond the inertial kernel approximation, which will also modify the noise spectrum at larger temperatures.

While we here discussed scalar couplings to the environment in depth, Eq.~\eqref{eq:modified_llg} does hold for any real 3x3 matrix $\Cwtens$ describing the spin-environment interaction in three dimensions.  Anisotropic coupling tensors can be implemented, suitable for describing magnetization dynamics within thin films~\cite{Chen2018}, where one direction is coupled differently to environmental modes than the other two.
One simplification of our three-dimensional model is to choose a coupling tensor such that  spins interact with only  one-dimensional environmental modes. This reduces the theory to the spin-boson model, see \Appendix~\ref{sec:spin-boson}, whose quantum thermodynamic properties have been discussed very extensively, recently for example in~\cite{Purkayastha2020}.

Microscopic heat transport in spin systems can also be analysed by allowing non--equilibrium situations where individual reservoir modes at  frequencies $\omega$ and for spins $n$  are thermal - but at different temperatures. This will result in spin dynamics that shuffles energy from one reservoir mode to another, and could result in two- and more-temperature models. For example, the possibility of different phonon modes, each with their own temperature, to couple with different strengths to electrons has recently been analysed in \cite{Maldonado2017} for a magnetic system excited by an ultra-short laser pulse.
Furthermore, in deriving the FDT we have assumed bosonic environmental modes but it would be insightful to identify changes to the properties of equation~\eqref{eq:modified_llg} that arise when the spins couple directly to electrons, or fermionic modes in general \cite{fermionic-Chen2013,fermionic-Strasberg2016,fermionic-Nazir}.

Beyond the quantum character of the reservoir, it will be important to numerically solve the full quantum dynamics according to Eq.~\eqref{eq:modified_llg}, including spin operators interacting with neighbouring spin operators. 
Advanced quantum numerical methods such as Hierarchical Equations Of Motion (HEOM)  \cite{Tanimurachapter2018}, and the recently proposed time-evolving matrix product operators (TEMPO) method  \cite{Strathearn2018} will be required to efficiently describe the time evolution of even just a single quantum spin coupled to a non-Markovian  environment.
For multiple interacting spins at low temperatures one can expect entanglement between the spins being present during the short-time dynamics, and even in steady state \cite{Nielsen1998, Arnesen2001,Zhang2005}. Unfortunately, evaluating such properties will very quickly become a numerically hard problem, requiring advanced numerical techniques such as density-matrix renormalisation group (DMRG)~\cite{Schollwoeck} to find realistic approximate solutions. 
Vice versa, solving \eqref{eq:modified_llg} within the classical spin vector approximation while including a full  quantum power spectrum for the environmental modes, may prove insightful and numerically tractable in the context of finding suitable models for noise in quantum computing hardware, such as superconducting qubits that are held in the $mK$ range~\cite{Dwave}. The results may also inform implementations of Young's double slit experiment with a levitated single magnetic domain nanoparticle using the Einstein-de Haas effect~\cite{Rusconi2017,Pino2018}. 

\smallskip 

\subsection*{Acknowledgments} 
\noindent We thank Karen Livesey, Richard Evans, Marco Berritta, Stefano Scali, Federico Cerisola, Luis Correa, James Cresser, Claudia Clarke, Ian Ford and Rob Hicken for inspiring discussions, Carsten Henkel and Richard Evans for comments on a draft of this manuscript, and Somayyeh Nemati for iron's Lorentzian parameters mentioned in \cite{Nemati2021}.  SARH thanks Paul Kinsler for pointing out the stupidity of numerically solving an integro--differential equation when an ODE will do.  SARH also acknowledges funding from the Royal Society and TATA (RPG-2016-186). CRJS and JA acknowledge support and funding from the EPSRC Centre for Doctoral Training in Electromagnetic Metamaterials EP/L015331/1.  JA acknowledges funding from EPSRC (EP/R045577/1) and the Royal Society.



\subsection*{Data  availability  statement} 
\noindent The Python code with which figures 2-6 were produced is available upon reasonable request to JA, janet@qipc.org.


\bigskip
\bigskip
\bigskip

\cleardoublepage

\appendix

\renewcommand{\thesubsection}{A\arabic{subsection}} 

\section*{Appendix} 

\subsection{Recovery of the spin-boson model} \label{sec:spin-boson}

The spin-boson model is recovered as a special case of the three-dimensional Hamiltonian \eqref{eq:total_hamiltonian}, and hence its dynamics is also given by Eq.~\eqref{eq:modified_llg}. To see this one may drop the site index $n$, and choose the external field as  
\be 
\Bext = B_0 \, (\cos (\theta) \, \boldsymbol{e}_z - \sin (\theta) \, \boldsymbol{e}_x),
\ee
for some angle $\theta$. Taking spin $1/2$ operators $\Spinwo=(\hbar/2)\boldsymbol{\sigma}$ and the coupling tensor as $(\Cwtens)_{jk} = \Cwwo \, \delta_{jk} \delta_{j1}$ with a scalar coupling function $\Cwwo$, one recovers from \eqref{eq:total_hamiltonian} the one-dimensional spin-boson Hamiltonian
\be 
    \H 
    &=& 	- \gamma B_0 \, (\cos (\theta) \,  \hat{S}_z - \sin (\theta) \, \hat{S}_x )   \nonumber \\
    && - \gamma \hat{S}_x \int_{0}^{\infty} \d\omega \, \, \Cwwo \,  \hat{x}_\omega  \\
    && + \frac{1}{2}  \int_{0}^{\infty} \d\omega \left[ \left(\hat{p}_{x , \omega}\right)^2 + \omega^{2}  \left(\hat{x}_\omega\right)^2 \right]  + \HR^{2D}, \nonumber
\ee 
where $\HR^{2D}$ is a decoupled two-dimensional reservoir that can be dropped from the dynamics. 

\subsection{Hermiticity of the quantum spin dynamics equation \label{ap:hermiticity}} 

The quantum spin dynamics equation \eqref{eq:modified_llg} is not written in an explicitly Hermitian form.  The integral term containing the damping kernel $\kernel{n}$ includes an operator product that does not equal its conjugate transpose
\begin{equation}
    \Spin{n}(t)\times\Spin{n}(t')\neq-\Spin{n}(t')\times\Spin{n}(t).
\end{equation}
Nevertheless equation \eqref{eq:modified_llg} is Hermitian, as it is the time integral that commutes with $\Spin{n}(t)$
\begin{equation}
    \left[\int_{\tin}^{t}\kernel{n}(t-t')\Spin{n}(t'),\Spin{n}(t)\right]=0\label{eq:kernel-commutes}
\end{equation}
This can be verified from an observation that Eq.~\eqref{eq:modified_llg} is simply a re--written form of the explicitly Hermitian equation \eqref{eq:spin_motion}.

Any confusion can be avoided through re--writing Eq.~\eqref{eq:modified_llg} in an equivalent but explicitly Hermitian form
\begin{multline}
    {\D\Spin{n}(t) \over \D t}= {\gamma \over 2} [\Spin{n}(t) \times \Beff{n}(t)\\
    - \Beff{n}(t) \times \Spin{n}(t)]\label{eq:explicitly-hermitian}
\end{multline}
where the effective magnetic field operator at time $t$ and site $n$ is given by
\begin{multline}
    \Beff{n}(t) =\Bext +{1 \over \gamma} \sum_{m\neq n} \bar{\mathcal{J}}^{(nm)}  \Spin{m} (t) +  \bstoch{n}(t) \\ + \gamma \int_{\tin}^{t} \d t' \, \kernel{n}(t-t')  \Spin{n} (t').
\end{multline}
In the Hermitian form (\ref{eq:explicitly-hermitian}) it is clear, for example that any term proportional to $\Spin{n}(t)$ appearing in $\Beff{n}$ does not affect the evolution of the spin operator, even though the operator cross product
\begin{equation}
    \Spin{n}(t)\times\Spin{n}(t)={\rm i}\hbar\Spin{n}(t)
\end{equation}
is non--zero.  A consequence of this result is that the zeroth order term in the expansion of the damping operator (\ref{eq:damping_expansion}) does not contribute to the evolution of the spin operator.

We note that Eq. (\ref{eq:kernel-commutes}) implies that only the sum of \textit{all} the terms in the damping kernel expansion (\ref{eq:damping_expansion}) commutes with the spin operator.  When using a truncated form of the expansion (\ref{eq:damping_expansion}) we must therefore use the explicitly Hermitian equation of motion (\ref{eq:explicitly-hermitian}).

\subsection{$\Spinwo^2$ is a constant of motion of Eq.~\eqref{eq:modified_llg}} \label{app:constantS} 

To evaluate the derivative of $(\Spin{n}(t))^2$ we first express Eq.~\eqref{eq:modified_llg} in explicitely Hermitian form \eqref{eq:explicitly-hermitian}.
Dropping site index and time for simplicity, we find
\begin{align}
\frac{\D |\Spinwo (t)|^2}{\D t} 
&= \sum_j \left(\hat{S}_j \frac{\D \hat{S}_j}{\D t} + \frac{\D \hat{S}_j}{\D t} \hat{S}_j \right) \nonumber \\
&= {\gamma \over 2}  \sum_{jkl}\epsilon_{jkl} \left( \hat{S}_j \hat{S}_k \hat{B}_l + \hat{B}_l \hat{S}_k\hat{S}_j \right) \nonumber  \\
&=\frac{{\rm i}\hbar\gamma}{2}\sum_{l}[\hat{S}_{l},\hat{B}_{l}]\nonumber\\
&=0,
\end{align}
where we have applied the angular momentum commutation relations, interchanged indices, and used the anti-symmetric property of $\epsilon_{jkm}$. The final line follows from the fact that the spin and the effective magnetic field commute.

\begin{figure*}[t]
\includegraphics[trim=0.8cm 0.8cm 0.42cm 0.38cm,clip, width=0.99\textwidth]{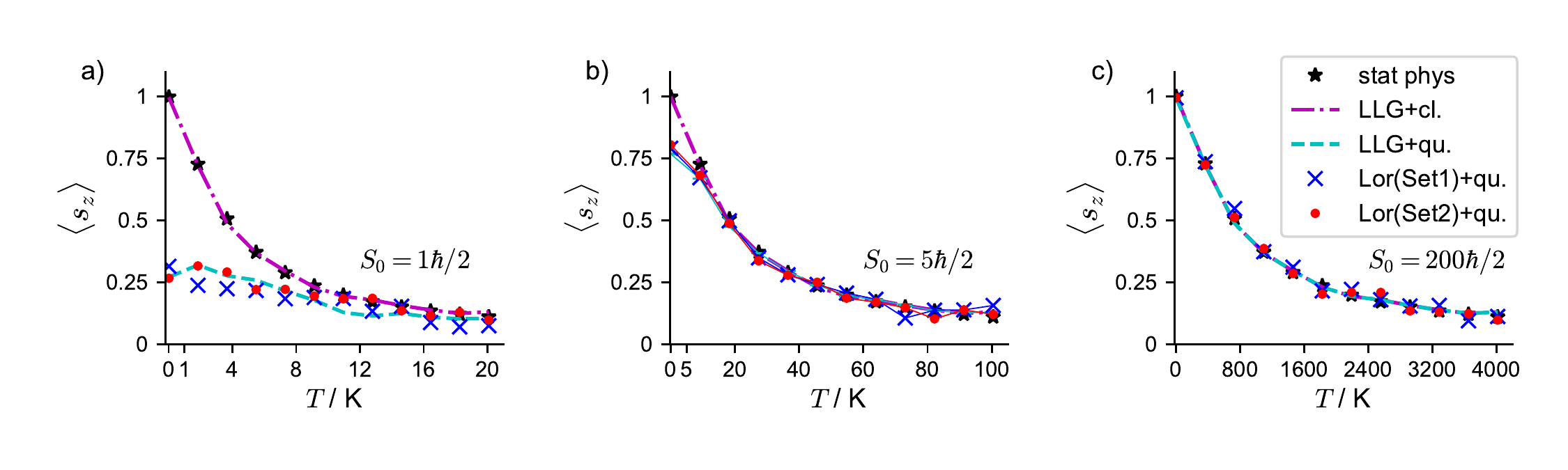}
\caption{ \label{fig:MoverT-appendix}  \sf 
Same notation and Lorentzian parameters as in Fig.~\ref{fig:MoverT} but here also showing spin length $S_0 = 5\hbar/2$ in panel b). Panels a) and c) show $\langle s_z\rangle$ for  $S_0 = 1\hbar/2$ and $S_0 = 200\hbar/2$, respectively, as in main text. Each curve is obtained by, for each temperature $T$, {\it time-averaging} over the late times of a single stochastic trajectory, from $0.75 \, t_{\rm max}$ to $t_{\rm max}=2 \pi \times 8000  \omega_L^{-1}$. 
}
\end{figure*}

\subsection{Lorentzian power spectrum expansion} \label{app:powerspecexpansion}

Similar to the damping kernel term expansion  \eqref{eq:damping_expansion}, in moments \eqref{eq:kappan_Lor} and time-derivatives, the Lorentzian power spectrum \eqref{eq:Lorpow} can be  expanded in powers of  frequency $\omega$, as
\begin{align}  \label{eq:Lorpow_expansion}
    \tilde{P}^\Lor_\quant (\omega) 
    =   \sum_{m=0}^{\infty}(-1)^{m+1}\omega^{2m+1}\kappa_{2m+1}^\Lor\coth\left(\frac{\hbar\omega}{2k_{B}T}\right),
\end{align}
where we have kept the quantum $\coth$ unexpanded.
The $\kappa^\Lor_{m}$ are the same  coefficients as those given in \eqref{eq:kappan_Lor}.  For small frequencies $\omega$ the first term in the series \eqref{eq:Lorpow_expansion} dominates and the power spectrum takes the (quantum) Ohmic form
\be
    \tilde{P}^{\rm Lor}_\quant(\omega)\approx-\omega \, \kappa_{1}^{\rm Lor}\coth\left(\frac{\hbar\omega}{2 k_{B}T}\right),
\ee
where comparison with \eqref{eq:llg_psd} again shows that $-\kappa_{1}^{\rm Lor}$ is the effective Gilbert damping constant. 

Beyond the Ohmic regime, one can see in \eqref{eq:Lorpow_expansion} that only the odd moments $\kappa^\Lor_{2m+1}$ contribute. Therefore the inertial term $\kappa_{2}^\Lor$, which is the first deviation of the damping kernel from Ohmic behaviour, does not change the quantum fluctuations in \eqref{eq:FDT}.  Only when the third order time derivative of the spin operator contributes significantly to equation \eqref{eq:modified_llg}, will memory effects begin to colour the spectrum away from the (quantum) Ohmic form \eqref{eq:llg_psd}.

\subsection{Set of equations for kernel simulation } \label{sec:kernel}

Here we show that the simulation of the kernel in Eq.~\eqref{eq:modified_llg} can be achieved by numerically integrating a set of first order coupled differential equations. We assume a single spin and rewrite Eq.~\eqref{eq:modified_llg}  as 
\be
	\frac{\D \clspin(t)}{\D t}= \gamma \clspin (t) \times 
	\bigg[  \Bext +  \boldsymbol{b}(t) + \boldsymbol{V} (t)  \bigg], \quad 
\ee
where we have defined
$\boldsymbol{V} (t)  =  \gamma \int_{\tin}^{t} \d t' \, K (t-t') \,  \clspin (t')$. Furthermore defining  $\boldsymbol{W} (t) =  \frac{\D \boldsymbol{V} (t)}{\D t}$, 
now leads to a differential equation for $\boldsymbol{W} (t)$:
\be
	 \frac{\D \boldsymbol{W} (t)}{\D t} 
	 &=&  \gamma \int_{\tin}^{t} \d t' \,  \frac{\D^2 K(t-t')}{\D t^2}  \,  \clspin (t'), 
\ee
where we have assumed $K(0)=0$ and $ \dot{K}(0)=0$. 
Expressing  $K(t-t')$ through its Fourier transform $\tilde{K} (\omega)$,
choosing a Lorentzian kernel~\eqref{eq:Lorkernel} and considering the expression $\boldsymbol{Z} (t) := \frac{\D \boldsymbol{W} (t)}{\D t} + \Gamma  \boldsymbol{W} (t) + \omega^2_0 \boldsymbol{V} (t)$, we obtain
\be
	\boldsymbol{Z} (t)
	 && = A \,  \gamma \, \int_{\tin}^{t} \d t' \,  \delta  (t-t') \,  \clspin (t')  .
\ee
Rearranging gives 
\be
	 \frac{\D \boldsymbol{W} (t)}{\D t} 
	 =  - \Gamma  \boldsymbol{W} (t) - \omega^2_0 \boldsymbol{V} (t) + A \, \gamma \,   \clspin (t)  , \quad \quad
\ee
as stated in the main text. (Note that the assumption $K(0)=0$ and $ \dot{K}(0)=0$ is fulfilled for the Lorentzian kernel, \eqref{eq:Lorkern}, since the Heaviside function $\Theta(\tau) = 1$ for $\tau>0$, and zero elsewhere.)

\subsection{Statistical physics prediction for $\langle s_z \rangle$ as function of temperature}\label{app:clMT}

For a classical spin $\boldsymbol{S}$ of length $S_0 = n \, {\hbar / 2}$ in an external field $\Bext = B \boldsymbol{e}_z$ the thermal average $\langle s_z \rangle$  is determined by the Boltzmann distribution for the Hamiltonian $H = -\gamma \, \boldsymbol{S} \cdot \Bext$ at inverse temperature $\beta =1/k_B T$,
\be
	\langle S_z \rangle_\stph 
	&=& \int_{- S_0}^{+ S_0} \d S_z \, S_z \,  {e^{- \beta (- \gamma \, S_z B)}  \over  Z_a  }
	= \partial_{a} \ln Z_a , \, \qquad 
\ee
with $Z_a  = \int_{- S_0}^{+ S_0} \, \, \d S_z \, \, e^{a \, S_z}$ where $a=\beta \gamma B$. This gives
\be
	Z_a   = {2 \sinh (a \, S_0) \over a} ,
\ee
and hence
\be
	{\langle S_z \rangle_\stph  \over  S_0}
	&=& \coth  \left({\beta \gamma B \, S_0} \right)  \, - {1 \over \beta \gamma B  \, S_0} , \\
	\langle s_z \rangle_\stph 
	&=& \coth  \left({n \, \hbar \omega_L   \over 2 k_B T} \right)  \, - {2 k_B T \over n \, \hbar \omega_L}, 
\ee
where $\omega_L = |\gamma B|$ and $s_z= \mbox{sign} (\gamma) \, {S_z \over  S_0}$. In the magnetism literature, sometimes a reduced temperature experienced by a spin with $n\neq1$ is defined as $T_{\rm red} = {T / n}$, i.e. the temperature is effectively reduced in comparison to the temperature experienced by a spin with $n=1$.

\subsection{Steady state $\langle s_z\rangle$ plot for spin $S_0 = 5\hbar/2$}\label{app:M(T)v2plot}

As discussed in the main text,  the impact of the quantum zero-point noise on the steady state $\langle s_z\rangle$ value at $T=0$K is very highly spin length dependent. 
For some materials a fundamental spin value of $S_0 = 1\hbar/2$ will not be appropriate. For example  Iron (III) has 5 electrons in the outer $d$ shell, and then from Hund's rules the spin is maximized to $S=5/2$, and the orbital angular momentum is zero, $L=0$.  Therefore $J=S$,  Land{\'e} g--factor equals 2, and the gyromagnetic ratio remains the electron gyromagnetic ratio. Fig.~\ref{fig:MoverT-appendix}b) shows the steady state $\langle s_z\rangle$ plot as a function of temperature for spin $S_0 = 5\hbar/2$, next to those for spin $S_0 = 1\hbar/2$ (a) and $S_0 =200\hbar/2$ (c).  The $\langle s_z \rangle$ value is below 1, at $\approx 0.8$, but the reduction is far less severe than for the spin-1/2.

\subsection{Scales in classical and for quantum thermostats}\label{app:scaling}

Here we establish the set of scales determining the dynamics described by Eqs.~\eqref{eq:llg_equation} and \eqref{eq:modified_llg} with either {\it classical} or {\it quantum} power spectra. 

For a single spin, i.e. ignoring exchange terms etc.,  one may rescale the LLG equation  \eqref{eq:llg_equation} using $\boldsymbol{M} =\gamma\boldsymbol{S}=|\gamma| S_0 \boldsymbol{s}$ with spin length $|\boldsymbol{S}| =S_0$. One obtains
\be	
    \frac{\D \boldsymbol{s}}{\D t}
	=\gamma\boldsymbol{s}\times\left[\Beffwo^\class (t) - |\gamma| S_0 \eta_G \frac{\D \boldsymbol{s}}{\D t}
	\right],
\ee
where the spin length $S_0$ and $\eta_G$ appear together, setting the first scale. 
Furthermore the effective field  including {\it classical} stochastic noise with power spectrum  \eqref{eq:ohmic_psd_classical}, is given through \eqref{eq:noisegeneration} and \eqref{eq:noisegeneration2} by 
\be 
    \Beffwo^\class (t) 
    =  \Bext +  \sqrt{ 2 S_0 \eta_G \, k_B {T \over S_0} } \, \boldsymbol{\xi}(t) .\label{eq:classfield}
\ee 
Here we have  introduced an $S_0$ so that $\eta_G$ appears together with it, and we find the second scale to be given by $T/S_0$.  The third scale is clearly set by the strength of the external field, $\Bext$. 
We note that if one chooses the same $S_0\eta_G$ value for different spin lengths, i.e. assumes $\eta_G$ scales as $1/S_0$, then only two scales are left,  $\Bext$ and $T/S_0$.

However, for the {\it quantum} Ohmic power spectrum the components of the stochastic noise can be written as
\begin{align} 
    b_j &=   \int_{-\infty}^{\infty} \d t' \, \int_{-\infty}^{\omega_c} {\, \d \omega \over 2\pi}  e^{- i \omega (t-t')} \nonumber \\
    & \times \sqrt{S_0 \eta_G  \, {\hbar \omega \over S_0} \coth\left(\frac{\hbar\omega}{2k_{B}T}\right)} \, \xi_j(t'), \label{eq:stochfield}
\end{align}
Clearly, in the {\it quantum} case the temperature $T$  now appears separately from spin length $S_0$, thus introducing an additional scale in comparison to the classical case. Moreover the fact that the frequency integration for the stochastic field does not simplify as in \eqref{eq:classfield} means that relaxation to the steady state at low temperatures (where the $\coth x$ cannot be approximated as $1/x$) will be much more noisy  than in the high temperature case. Thus in our simulations, this additional scale leads to larger uncertainties in the steady state results, as seen in Fig.~\ref{fig:MoverT}a).

Finally, we note that for the integration of the quantum Ohmic power spectrum in \eqref{eq:stochfield} we have introduced a  frequency cut-off $\omega_c$ by hand, which is necessary at low temperatures to avoid the integral diverging. At low temperatures, this cut-off will set an additional, somewhat artificial, scale of the problem. Importantly, such cut-off is not required for the Lorentzian coupling since the power spectrum \eqref{eq:Lorpow} decays at high frequencies, even at low $T$.


\begin{thebibliography}{12}%
\makeatletter
\providecommand \@ifxundefined [1]{%
 \@ifx{#1\undefined}
}%
\providecommand \@ifnum [1]{%
 \ifnum #1\expandafter \@firstoftwo
 \else \expandafter \@secondoftwo
 \fi
}%
\providecommand \@ifx [1]{%
 \ifx #1\expandafter \@firstoftwo
 \else \expandafter \@secondoftwo
 \fi
}%
\providecommand \natexlab [1]{#1}%
\providecommand \enquote  [1]{``#1''}%
\providecommand \bibnamefont  [1]{#1}%
\providecommand \bibfnamefont [1]{#1}%
\providecommand \citenamefont [1]{#1}%
\providecommand \href@noop [0]{\@secondoftwo}%
\providecommand \href [0]{\begingroup \@sanitize@url \@href}%
\providecommand \@href[1]{\@@startlink{#1}\@@href}%
\providecommand \@@href[1]{\endgroup#1\@@endlink}%
\providecommand \@sanitize@url [0]{\catcode `\\12\catcode `\$12\catcode
  `\&12\catcode `\#12\catcode `\^12\catcode `\_12\catcode `\%12\relax}%
\providecommand \@@startlink[1]{}%
\providecommand \@@endlink[0]{}%
\providecommand \url  [0]{\begingroup\@sanitize@url \@url }%
\providecommand \@url [1]{\endgroup\@href {#1}{\urlprefix }}%
\providecommand \urlprefix  [0]{URL }%
\providecommand \Eprint [0]{\href }%
\providecommand \doibase [0]{http://dx.doi.org/}%
\providecommand \selectlanguage [0]{\@gobble}%
\providecommand \bibinfo  [0]{\@secondoftwo}%
\providecommand \bibfield  [0]{\@secondoftwo}%
\providecommand \translation [1]{[#1]}%
\providecommand \BibitemOpen [0]{}%
\providecommand \bibitemStop [0]{}%
\providecommand \bibitemNoStop [0]{.\EOS\space}%
\providecommand \EOS [0]{\spacefactor3000\relax}%
\providecommand \BibitemShut  [1]{\csname bibitem#1\endcsname}%
\let\auto@bib@innerbib\@empty
\bibitem [{Note1()}]{Note1}%
  \BibitemOpen
  \bibinfo {note} {Note that Eq.~\protect \textup {\hbox {\mathsurround \z@
  \protect \normalfont (\ignorespaces \ref {eq:llg_equation}\unskip
  \@@italiccorr )}} is expressed in SI rather than Gaussian units (as in
  e.g.~\cite {lakshmanan2011}).}\BibitemShut {Stop}%
\bibitem [{Note2()}]{Note2}%
  \BibitemOpen
  \bibinfo {note} {Not the Brownian motion Brown!}\BibitemShut {Stop}%
\bibitem [{Note3()}]{Note3}%
  \BibitemOpen
  \bibinfo {note} {Using vector identities the time derivative on the right
  hand side of Eq.~\protect \textup {\hbox {\mathsurround \z@ \protect
  \normalfont (\ignorespaces \ref {eq:llg_equation}\unskip \@@italiccorr )}}
  can be eliminated and the equation becomes $\partial \protect \boldsymbol {M}
  / \partial t=\gamma '\protect \boldsymbol {M}\times \protect \boldsymbol
  {B}_{\protect \rm eff}-\lambda \protect \boldsymbol {M}\times \left (\protect
  \boldsymbol {M}\times \protect \boldsymbol {B}_{\protect \rm eff} \right )$
  with $\gamma '$ and $\lambda $ functions of $\gamma , \eta _G$ and $|\protect
  \boldsymbol {M}|$.}\BibitemShut {Stop}%
\bibitem [{Note4()}]{Note4}%
  \BibitemOpen
  \bibinfo {note} {We use SI rather than Gaussian units, so we have $B$ (units
  ${\protect \rm mass}/({\protect \rm charge}\times {\protect \rm time})$)
  rather than an $H$-field (units ${\protect \rm charge}/({\protect \rm
  length}\times {\protect \rm time})$).}\BibitemShut {Stop}%
\bibitem [{Note5()}]{Note5}%
  \BibitemOpen
  \bibinfo {note} {Here the spins are discrete and positioned on a lattice, but
  one could also use a continuum description, as in micromagnetics~\cite
  {mayergoyz2009}.}\BibitemShut {Stop}%
\bibitem [{Note6()}]{Note6}%
  \BibitemOpen
  \bibinfo {note} {Note that tensors and vectors are set in calligraphic and
  bold font, respectively, and that scalar products between vectors are
  indicated with $\cdot $, while a tensor followed by a tensor or a vector is
  to be understood as matrix multiplication.}\BibitemShut {Stop}%
\bibitem [{Note7()}]{Note7}%
  \BibitemOpen
  \bibinfo {note} {Note that in contrast to what is typically done in
  Caldeira--Leggett type models\cite {caldeira} no counter term has been
  included here. In any case, coupling to a spin would result in a term
  proportional to $\protect \boldsymbol {S}^2 \propto \protect \mathbbm {1}$,
  which will incur an offset in the overall Hamiltonian that does not affect
  the dynamics.}\BibitemShut {Stop}%
\bibitem [{Note8()}]{Note8}%
  \BibitemOpen
  \bibinfo {note} {The Fourier transform $\protect \mathaccentV
  {tilde}07E{{\protect \mathcal K}}^{(n)}(\omega )$ of the kernel ${\protect
  \mathcal K}^{(n)}\protect \tmspace -\thinmuskip {.1667em}(\tau )$
  automatically satisfies the Kramers--Kronig relations~\cite {volume5},
  connecting the dissipative and reactive parts of the response kernel, as is
  required for any causal response.}\BibitemShut {Stop}%
\bibitem [{Note9()}]{Note9}%
  \BibitemOpen
  \bibinfo {note} {The autocorrelation function of two reservoir operators $A$
  and $B^\protect \dag $ in the thermal reservoir state $\protect \mathaccentV
  {hat}05E{\rho }_R$ is defined as the expectation value of the Hermitian
  operator, ${\delimiter "426830A \left \protect \{A(t) , B^{\protect \dag
  }(t-\tau ) \right \protect \} \delimiter "526930B _{\beta }/2}$. In the
  classical case $A$ and $B^\protect \dag $ commute at all times, removing the
  need for this distinction.}\BibitemShut {Stop}%
\bibitem [{Note10()}]{Note10}%
  \BibitemOpen
  \bibinfo {note} {The Fourier transform is here defined as $\protect
  \mathaccentV {tilde}07E{f} (\omega ) = \DOTSI \intop \ilimits@ _{-\infty
  }^{\infty } \protect \tmspace -\thinmuskip {.1667em}\protect \tmspace
  -\thinmuskip {.1667em}\protect \tmspace -\thinmuskip {.1667em}\protect
  \mathrm {d}\tau \protect \tmspace +\thinmuskip {.1667em} e^{+{\protect \rm i}
  \omega \tau } f(\tau )$, with the inverse $f(\tau ) = \DOTSI \intop \ilimits@
  _{-\infty }^{\infty } {\protect \tmspace -\thinmuskip {.1667em}\protect
  \tmspace -\thinmuskip {.1667em}\protect \tmspace -\thinmuskip
  {.1667em}\protect \mathrm {d}\omega \over 2 \pi } \protect \tmspace
  +\thinmuskip {.1667em} e^{-{\protect \rm i} \omega \tau } \protect
  \mathaccentV {tilde}07E{f} (\omega )$.}\BibitemShut {Stop}%
\bibitem [{Note11()}]{Note11}%
  \BibitemOpen
  \bibinfo {note} {The power spectrum given in \protect \textup {\hbox
  {\mathsurround \z@ \protect \normalfont (\ignorespaces \ref {eq:FDT}\unskip
  \@@italiccorr )}} is the correct general version for any kernel ${\protect
  \mathcal K}^{(n)}\protect \tmspace -\thinmuskip {.1667em}(t)$, fulfilling the
  full quantum FDT~\cite {volume5}. For a Gilbert damping kernel a power
  spectrum proportional to $\hbar \omega /(\protect \qopname \relax
  o{exp}{(\hbar \omega /k_B T)}-1)$ was given in~\cite
  {Oppeneer1998,woo2015,bergqvist2018,Barker2019}. This is missing the quantum
  ground state contribution of $\hbar \omega /2$, which acts as stochastic
  noise on the spin system even at zero temperature.}\BibitemShut {Stop}%
\bibitem [{Note12()}]{Note12}%
  \BibitemOpen
  \bibinfo {note} {Note that although \protect \textup {\hbox {\mathsurround
  \z@ \protect \normalfont (\ignorespaces \ref {eq:LLGdamping}\unskip
  \@@italiccorr )}} is not an explicitly Hermitian operator, Appendix~\ref
  {ap:hermiticity} shows that Eq.~\protect \textup {\hbox {\mathsurround \z@
  \protect \normalfont (\ignorespaces \ref {eq:modified_llg}\unskip
  \@@italiccorr )}} is nevertheless equivalent to a Hermitian equation of
  motion.}\BibitemShut {Stop}%
\end{thebibliography}%


%


\begin{thebibliography}{99}

\bibitem{Goold2016}
J. Goold, M. Huber, A. Riera, L. del Rio, and P. Skrzypczyk,
``The role of quantum information in thermodynamics -- a topical review'',
{\it J. Phys. A: Math. Theor.} \textbf{49}, 143001 (2016).

\bibitem{Vinjanampathy2016}
S. Vinjanampathy, J. Anders,
``Quantum thermodynamics'',
{\it Contemp. Phys.} \textbf{57}, 545 (2016).

\bibitem{Book2018}
{\it Thermodynamics in the Quantum Regime: Fundamental Aspects and New Directions}, edited by F. Binder {\it et al.}, 
Springer, (2018). 

\bibitem{Wichterich2007}
H. Wichterich, M.J. Henrich, H.-P. Breuer, J. Gemmer,  M. Michel,
``Modeling heat transport through completely positive maps'',
{\it Phys. Rev. E} \textbf{76}, 031115 (2007).

\bibitem{Boudjada2014}
N. Boudjada, and  D.  Segal,
``From Dissipative Dynamics to Studies of Heat Transfer at the Nanoscale: Analysis of the Spin-Boson Model'',
{\it J. Phys.  Chem.  A} \textbf{118}, 11323 (2014).

\bibitem{Yang2014}
Y. Yang and C.Q. Wu, 
``Quantum heat transport in a spin-boson nanojunction: Coherent and incoherent mechanisms'',
{\it Europhys. Lett.} \textbf{107}, 30003 (2014).

\bibitem{Freitas2017}
N. Freitas, J.P. Paz,
``Fundamental limits for cooling of linear quantum refrigerators'',
{\it Phys. Rev. E} \textbf{95}, 012146  (2017).

\bibitem{Funo2018}
K. Funo and H. T. Quan,
``Path integral approach to heat in quantum thermodynamics'',
{\it Phys. Rev. E} \textbf{98}, 012113 (2018).

\bibitem{Whitney2018}
R.S. Whitney, R. Sanchez, J. Splettstoesser,
``Quantum thermodynamics of nanoscale thermoelectrics and electronic devices'',
in {\it Thermodynamics in the Quantum Regime}
Springer, (2018). 

\bibitem{Yang2019} 
J. Yang, C. Elouard, J. Splettstoesser, B. Sothmann, R. Sanchez,  A.N. Jordan,
``Thermal transistor and thermometer based on Coulomb-coupled conductors'',
{\it Phys. Rev. B} \textbf{100}, 045418 (2019).

\bibitem{Benatti2020}
F. Benatti, R. Floreanini and L. Memarzadeh,
``Bath-assisted transport in a three-site spin chain: Global versus local approach'',
{\it Phys. Rev. A} {\bf 102}, 042219 (2020).



\bibitem{Maniscalco}
S. Maniscalco and F Petruccione,
``Non-Markovian dynamics of a qubit'',
{\it Phys. Rev. A} \textbf{73}, 012111  (2006).


\bibitem{Rivas2010}
A. Rivas, A.D.K. Plato, S.F. Huelga and M.B. Plenio, 
``Markovian master equations: a critical study'',
{\it New J. Phys.} {\bf 12}, 113032 (2010).


\bibitem{fermionic-Chen2013}
M. Chen and J. Q. You,
``Non-Markovian quantum state diffusion for an open quantum system in fermionic environments'',
{\it Phys. Rev. A} \textbf{87}, 052108 (2013).

\bibitem{fermionic-Strasberg2016}
Ph. Strasberg, G. Schaller, N. Lambert and T. Brandes,
``Nonequilibrium thermodynamics in the strong coupling and non-Markovian regime based on a reaction coordinate mapping'',
{\it New J. Phys.} {\bf 18}, 073007 (2016).

\bibitem{deVega}
I. de Vega and D. Alonso,
``Dynamics of non-Markovian open quantum systems'',
{\it Rev. Mod. Phys.} \textbf{89}, 015001 (2017).

\bibitem{Cianciaruso2017}
M. Cianciaruso, S. Maniscalco, and G. Adesso,
``Role of non-Markovianity and backflow of information in the speed of quantum evolution'',
{\it Phys. Rev. A} \textbf{96}, 012105 (2017).

\bibitem{Strasberg2018}
Ph. Strasberg, M. Esposito,                           
``Response Functions as Quantifiers of Non-Markovianity'',
{\it Phys. Rev. Lett.} \textbf{121}, 040601 (2018).

\bibitem{Raja2018}
S.H.Raja, M. Borrelli, R. Schmidt, J.P. Pekola, and S. Maniscalco,
``Thermodynamic fingerprints of non-Markovianity in a system of coupled superconducting qubits'',
{\it Phys. Rev. A} \textbf{97}, 032133 (2018).

\bibitem{Uzdin2015}
R. Uzdin, A. Levy, R. Kosloff,
``Equivalence of quantum heat machines, and quantum-thermodynamic signatures'',
{\it Phys. Rev. X}\textbf{5}  031044 (2015).

\bibitem{Brask2015}
J. Bohr Brask, G. Haack, N. Brunner, M. Huber,
``Autonomous quantum thermal machine for generating steady-state entanglement'',
{\it New J. Phys.}\textbf{17}  113029 (2015).

\bibitem{Kammerlander2016}
P. Kammerlander, J. Anders,
``Coherence and measurement in quantum thermodynamics'',
{\it Sci. Rep.} \textbf{6}  1 (2016).

\bibitem{Sapienza2019}
F. Sapienza, F. Cerisola, A. J. Roncaglia,
``Correlations as a resource in quantum thermodynamics'',
{\it Nat. Comms} \textbf{10} 1 (2019).

\bibitem{Klatzow2019}
J. Klatzow, J.N. Becker, P.M. Ledingham, {\it et al.},  
``Experimental demonstration of quantum effects in the operation of microscopic heat engines'',
{\it Phys. Rev. Lett.} \textbf{122} 110601 (2019).

\bibitem{Seifert2016}
U. Seifert,          
``First and Second Law of Thermodynamics at Strong Coupling'',
{\it Phys. Rev. Lett.} \textbf{116} 020601 (2016).

\bibitem{philbinanders2016} 
T. G. Philbin and J. Anders, 
``Thermal energies of classical and quantum damped oscillators coupled to reservoirs'', 
{\it J. Phys. A} \textbf{49}, 215303 (2016).

\bibitem{Jarzynski2017} 
C. Jarzynski,  
``Stochastic and Macroscopic Thermodynamics of Strongly Coupled Systems'',
{\it Phys. Rev. X} \textbf{7} 011008 (2017).

\bibitem{Miller2017}
H. Miller, J. Anders, 
``Entropy production and time asymmetry in the presence of strong interactions'',
{\it Phys. Rev. E} \textbf{95} 062123 (2017).

\bibitem{Cresser2017}
J. D. Cresser, C. Facer
``Coarse-graining in the derivation of Markovian master equations and its significance in quantum thermodynamics'',
arXiv:1710.09939  (2017).

\bibitem{Miller2019}
H. J. D. Miller and J. Anders, 
``Energy-temperature uncertainty relation in quantum thermodynamics'',
{\it Nat. Comms.} \textbf{9}, 2203 (2018).

\bibitem{Kawai2019}
R. Kawai, K. Goyal,
``Steady state thermodynamics of two qubits strongly coupled to bosonic environments'',
{\it Phys. Rev. Res.} \textbf{1}, 033018 (2019).

\bibitem{Strasberg2020}
Ph. Strasberg and M. Esposito
``Measurability of nonequilibrium thermodynamics in terms of the Hamiltonian of mean force''
{\it Phys. Rev. E} \textbf{101}, 050101(R) (2020).

\bibitem{Purkayastha2020}
A. Purkayastha, G. Guarnieri, M. Mitchison, R. Filip, J. Goold, 
``Tunable phonon-induced steady-state coherence in a double-quantum-dot charge qubit'',
{\it npj Quant. Inf.} \textbf{6} 27 (2020).


\bibitem{Kenawy2018}
A. Kenawy, J. Splettstoesser and M. Misiorny,
``Vibration-induced modulation of magnetic anisotropy in a magnetic molecule'',
{\it Phys. Rev. B} {\bf 97}, 235441 (2018).




\bibitem{caldeira} 
A. O. Caldeira and A. J. Leggett, 
``Path integral approach to quantum Brownian motion'', {\it Physica A} \textbf{121}, 587 (1983).

\bibitem{Hu92}
B.L. Hu, J.P. Paz and Y. Zhang,
``Quantum Brownian motion in a general environment: Exact master equation with nonlocal dissipation and colored noise'',
{\it Phys. Rev. D} \textbf{45}, 2843 (1992).

\bibitem{Thoss2001}
M. Thoss, H. Wang, W. H. Miller,
``Self-consistent hybrid approach for complex systems: Application to the spin-boson model with Debye spectral density'',
{\it J. Chem. Phys.}, \textbf{115} 2991 (2001).

\bibitem{Breuer-Petruccione}
H.P. Breuer, F. Petruccione,
``The Theory of Open Quantum Systems'',
Oxford University Press, (2002).

\bibitem{ABV2007}
F. B. Anders, R. Bulla, and M. Vojta,
``Equilibrium and Nonequilibrium Dynamics of the Sub-Ohmic Spin-Boson Model'',
{\it Phys. Rev. Lett.} \textbf{98}, 210402 (2007).

\bibitem{Huelga2013}
S.F. Huelga and M.B. Plenio,
``Vibrations, quanta and biology'',
{\it Cont. Phys.} \textbf{54} 181 (2013).

\bibitem{Nazir2016}
A. Nazir and D. P. S. McCutcheon,
``Modelling exciton'' phonon interactions in optically driven quantum dots'',
{\it Jour. of Phys.: Cond. Mat.} \textbf{28} 10 (2016). 

\bibitem{Seagate}
\href{https://www.seagate.com/gb/en/innovation/hamr/}{https://www.seagate.com/gb/en/innovation/hamr/}, accessed June 2021.

\bibitem{gilbert}
T. L. Gilbert,
``A Lagrangian formulation of the gyromagnetic equation of the magnetic field'',
{\it Phys. Rev.} \textbf{100}, 1243 (1955);
and 
T. L. Gilbert, 
``A phenomenological theory of damping in ferromagnetic materials'', 
{\it IEEE Trans. Mag.} \textbf{40}, 3443 (2004).

\bibitem{mayergoyz2009} 
I. D. Mayergoyz, G. Bertotti, and C. Serpico, ``Nonlinear Magnetization Dynamics in Nanosystems'', Elsevier (2009).

\bibitem{lakshmanan2011} 
M. Lakshmanan, 
``The fascinating world of the Landau--Lifshitz--Gilbert equation: an overview'', 
{\it Phil. Trans. R. Soc. A} \textbf{369}, 1280 (2011).

\bibitem{vansteenkiste2014} 
A. Vansteenkiste,  J. Leliaert, M. Dvornik, M. Helsen, F. Garcia--Sanchez and B. V. Waeyenberge, 
``The design and verification of MuMax3'', 
{\it AIP Adv.} \textbf{4}, 107133 (2014).

\bibitem{evans2014} 
R. F. L. Evans, W. J. Fan, P. Chureemart, T. A. Ostler, M. O. A. Ellis and R. W. Chantrell, 
``Atomistic spin model simulations of magnetic nanomaterials'', 
{\it J. Phys.: Condens. Matter} \textbf{26} 103202 (2014).

\bibitem{brown1963} 
W. F. Brown,
``Thermal Fluctuations of a Single-Domain Particle'', 
{\it Phys. Rev.} \textbf{130}, 1677 (1963).

\bibitem{ciornei2011} 
M.-C. Ciornei, J. M. Rub\'i, and J.-E. Wegrowe, 
``Magnetization dynamics in the inertial regime: Nutation predicted at short time scales'', 
{\it Phys. Rev. B} \textbf{83}, 020410(R) (2011).

\bibitem{Neeraj2021}
K. Neeraj, N.  Awari, S. Kovalev,  {\it et al.},  
``Inertial spin dynamics in ferromagnets'',
{\it Nat. Phys.} \textbf{17}, 245 (2021).

\bibitem{rebei2003} 
A. Rebei and G. J. Parker, 
``Fluctuations and dissipation of coherent magnetization'', 
{\it Phys. Rev. B} \textbf{67}, 104434 (2003).

\bibitem{garcia-palacios1999} 
J. L. Garcia-Palacios, 
``Brownian rotation of classical spins: dynamical equations for non-bilinear spin-environment couplings'', 
{\it Eur. Phys. J. B} \textbf{11}, 293 (1999).

\bibitem{Rossi2005} 
E. Rossi, O.G. Heinonen and A. H. MacDonald, 
``Dynamics of magnetization coupled to a thermal bath of elastic modes'', 
{\it Phys. Rev. B} \textbf{72}, 174412 (2005).

\bibitem{Bauer08} 
A. Brataas, Y. Tserkovnyak, and G.E.W. Bauer,
``Scattering Theory of Gilbert Damping'',
{\it Phys. Rev. Lett.} {\bf 101}, 037207 (2008).

\bibitem{bose2011} 
T. Bose and S. Trimpert, 
``Retardation effects in the Landau-Lifshitz-Gilbert equation'', 
{\it Phys. Rev. B} \textbf{83}, 134434 (2011).

\bibitem{schutte2014}
C. Sch\"utte, J. Iwasaki, A. Rosch, and N. Nagaosa, 
``Inertia, diffusion, and dynamics of a driven skyrmion'', 
{\it Phys. Rev. B} \textbf{90}, 174434 (2014).

\bibitem{thonig2015} 
D. Thonig, J. Henk, and O. Eriksson, 
``Gilbert-like damping caused by time retardation in atomistic magnetization dynamics'', 
{\it Phys. Rev. B} \textbf{92}, 104403 (2015).

\bibitem{bajpai2019}
U. Bajpai and B. K. Nikolic, 
``Time-retarded damping and magnetic inertia in the Landau-Lifshitz-Gilbert equation self-consistently coupled to electronic time-dependent nonequilibrium Green functions'', 
{\it Phys. Rev. B} \textbf{99}, 134409 (2019).

\bibitem{li2015} 
Y. Li, A.-L. Barra, S. Auffret, U. Ebels, and W. E. Bailey, 
``Inertial terms to magnetization dynamics in ferromagnetic thin films'', 
{\it Phys. Rev. B}  \textbf{92}, 140413(R) (2015).

\bibitem{Barker2019} 
J. Barker and G. E. W. Bauer, 
``Semiquantum thermodynamics of complex ferrimagnets'', 
{\it Phys. Rev. B} \textbf{100}, 140401(R) (2019).

\bibitem{beaurepaire1996} 
E. Beaurepaire, J.-C. Merle, A. Daunois, and J.-Y. Bigot, 
``Ultrafast Spin Dynamics in Ferromagnetic Nickel'', 
{\it Phys. Rev. Lett.} \textbf{76}, 4250 (1996).

\bibitem{Chen2018}
L. Chen, S. Mankovsky, S. Wimmer, {\it et al.},  
``Emergence of anisotropic Gilbert damping in ultrathin Fe layers on GaAs(001)'', 
{\it Nat. Phys.} \textbf{14}, 490 (2018). 

\bibitem{volume5} 
L. D. Landau and E. M. Lifshitz, 
``Statistical Physics (Part 1)'', Butterworth--Heinemann (2005).

\bibitem{huttner1992} 
B. Huttner and S. M. Barnett, 
``Quantization of the electromagnetic field in dielectrics'', 
{\it Phys. Rev. A} \textbf{46}, 4306 (1992).

\bibitem{azzawi2017} 
S. Azzawi, A. T. Hindmarch, and D. Atkinson, 
``Magnetic damping phenomena in ferromagnetic thin-films and multilayers'', 
{\it J. Phys. D: Appl. Phys.} \textbf{50}, 473001 (2017).


\bibitem{scheel2008} 
S. Scheel and S. Y. Buhmann, 
``Macroscopic QED - concepts and applications'', 
{\it Acta Physica Slovaca} \textbf{58}, 675 (2008).

\bibitem{philbin2010} 
T. G. Philbin, 
``Canonical quantization of macroscopic electromagnetism'', 
{\it New J. Phys.} \textbf{12}, 123008 (2010).

\bibitem{philbin2011}
T. G. Philbin, 
``Casimir effect from macroscopic quantum electrodynamics'', 
{\it New J. Phys.} \textbf{13}, 063026 (2011).

\bibitem{nieves2014}
P. Nieves, D. Serantes, U. Atxitia, and O. Chubykalo-Fesenko,
``Quantum Landau-Lifshitz-Bloch equation and its comparison with the classical case'',
{\it Phys. Rev. B} \textbf{90}, 104428 (2014).

\bibitem{Atxitia2009} 
U. Atxitia, O. Chubykalo-Fesenko, R. W. Chantrell, U. Nowak, and A. Rebei,
``Ultrafast Spin Dynamics: The Effect of Colored Noise'',
{\it Phys. Rev. Lett.} \textbf{102}, 057203 (2009).

\bibitem{Strungaru2021} 
M. Strungaru, M.O.A. Ellis, S. Ruta, O. Chubykalo-Fesenko, R.F.L. Evans, and R.W. Chantrell,
``Spin-lattice dynamics model with angular momentum transfer for canonical and microcanonical ensembles'',
{\it Phys. Rev. B} \textbf{103}, 024429 (2021).

\bibitem{asmann2019} 
M. Asmann and U. Nowak, 
``Spin-lattice relaxation beyond Gilbert damping'',
{\it J. Mag. and Mag. Mat.} \textbf{469}, 217 (2019).

\bibitem{fahnle2019} 
M. F\"ahnle, 
``Comparison of theories of fast and ultrafast magnetization dynamics'', 
{\it J. Mag. and Mag. Mat.} \textbf{469}, 28 (2019).

\bibitem{Oppeneer1998}
S.V. Halilov, H. Eschrig, A.Y. Perlov, and P.M. Oppeneer,
``Adiabatic spin dynamics from spin-density-functional theory: Application to Fe, Co, and Ni'',
{\it Phys. Rev. B} \textbf{58}, 293 (1998).

\bibitem{woo2015}
C.H. Woo, H. Wen, A.A. Semenov, S.L. Dudarev,  and P.W. Ma, 
``Quantum heat bath for spin-lattice dynamics'', 
{\it Phys. Rev. B} \textbf{91}, 104306 (2015).

\bibitem{bergqvist2018} 
L. Bergqvist and A. Bergman, 
``Realistic finite temperature simulations of magnetic systems using quantum statistics'', 
{\it Phys. Rev. Mat.} \textbf{2}, 013802 (2018).

\bibitem{Barker2020}
J. Barker, D. Pashov, J. Jackson, 
``Electronic structure and finite temperature magnetism of yttrium iron garnet'',
{\it Electron. Struct.} {\bf 2}, 044002 (2020).

\bibitem{Correa2019}
L.A. Correa, B. Xu, B. Morris, and G. Adesso,
``Pushing the limits of the reaction-coordinate mapping'',
{\it J. Chem. Phys.} \textbf{151}, 094107 (2019).

\bibitem{Nemati2021}
S. Nemati, C. Henkel, J. Anders, et al., in preparation (summer 2021). 
\quad For example for the phonon environment at 300K in bcc $\alpha$-57Fe \cite{Mauger2014}, these parameters are of the order of $\omega_0 \approx 2\pi * \, 6$ THz and $\Gamma \approx 2\pi * \, 3.5$ THz.

\bibitem{Mauger2014}
L. Mauger, M.S. Lucas, J.A. Munoz,  S.J. Tracy, M. Kresch,  Y. Xiao, P. Chow,  and B. Fultz,
``Nonharmonic phonons in $\alpha$-iron at high temperatures'',
{\it Phys. Rev. B} \textbf{90}, 064303 (2014).

\bibitem{Miller2018}
H. Miller, 
``Hamiltonian of mean force for strongly-coupled systems'', 
in {\it Thermodynamics in the Quantum Regime}
Springer, (2018). 

\bibitem{lu2012}
J.-T. L{\"u}, M. Brandbyge, P. Hedeg{\r{a}}rd, T. N. Todorov, and D. Dundas, 
``Current-induced atomic dynamics, instabilities, and Raman signals: Quasiclassical Langevin equation approach'', 
{\it Phys. Rev. B}, \textbf{85}, 245444 (2012).

\bibitem{lu2019}
J.-T. L{\"u}, B.-Z. Hu, P. Hedeg{\r{a}}rd and M. Brandbyge,
``Semi-classical generalized Langevin equation for equilibrium and nonequilibrium molecular dynamics simulation'',
{\it Prog. Surf. Sci.}, \textbf{94}, 21 (2019).

\bibitem{koch1980}
R. H. Koch, D. J. Van Harlingen and J. Clarke,
``Quantum-noise theory for the resistively shunted Josephson junction'', 
{\it Phys. Rev. Lett.}, \textbf{45}, 2132 (1980).

\bibitem{schmid1982}
A. Schmid, 
``On a quasiclassical Langevin equation'', 
{\it J. Low Temp. Phys.} \textbf{49}, 609 (1982).

\bibitem{kleinert1995}
H. Kleinert and S.V. Shabanov,
``Quantum Langevin equation from forward-backward path integral'',
{\it Phys. Lett. A}, \textbf{200}, 224 (1995).

\bibitem{eckern1990}
U. Eckern, W. Lehr, A. Menzel-Dorwarth, F. Pelzer, and A. Schmid, ``The quasiclassical Langevin equation and its application to the decay of a metastable state and to quantum fluctuations'', 
{\it J. Stat. Phys.} \textbf{59}, 885 (1990).

\bibitem{SchmidtMeistrenko2015} 
J. Schmidt, A.Meistrenko, H. van Hees, Z. Xu, and C. Greiner, ``Simulation of stationary Gaussian noise with regard to the Langevin equation with memory effect'', 
{\it Phys. Rev. E} \textbf{91}, 032125 (2015).	

\bibitem{scipy} 
P. Virtanen, {\it et al.}, 
``SciPy 1.0: fundamental algorithms for scientific computing in Python'', 
{\it Nat. Meth.} \textbf{17}, 261 (2020).

\bibitem{Evans15} 
R.F.L. Evans, U. Atxitia, and R.W. Chantrell,
``Quantitative simulation of temperature-dependent magnetization dynamics and equilibrium properties of elemental ferromagnets'',
{\it Phys. Rev. B} \textbf{91}, 144425 (2015).

\bibitem{Kuzmin}
M. D. Kuz' min, 
``Shape of Temperature Dependence of Spontaneous Magnetization of Ferromagnets: Quantitative Analysis'',
{\it Phys. Rev. Lett.} \textbf{94}, 107204 (2005).

\bibitem{Kuhn2017}
S. Kuhn, A. Kosloff, B. A. Stickler, F. Patolsky, K. Hornberger, M. Arndt, and James Millen, 
``Full rotational control of levitated silicon nanorods'', 
{\it Optica} \textbf{4}, 356 (2017).

\bibitem{Stickler2018}
B. A. Stickler, B. Schrinski, and K. Hornberger,
``Rotational Friction and Diffusion of Quantum Rotors'',
{\it Phys. Rev. Lett.} \textbf{121}, 040401 (2018).

\bibitem{Kusminskiy2016}
S. V. Kusminskiy, H. X. Tang, and F. Marquardt,
``Coupled spin-light dynamics in cavity optomagnonics''
{\it Phys. Rev. A} {\bf 94}, 033821 (2016).

\bibitem{Maldonado2017}
P. Maldonado, K. Carva, M. Flammer, and P.M. Oppeneer, 
``Theory of out-of-equilibrium ultrafast relaxation dynamics in metals'', 
{\it Phys. Rev. B} \textbf{96}, 174439 (2017).

\bibitem{fermionic-Nazir}
C. McConnell and A. Nazir,
``Electron counting statistics for non-additive environments'',
{\it J. Chem. Phys.} \textbf{151}, 054104 (2019). 

\bibitem{Tanimurachapter2018}
A. Kato, and Y. Tanimura,
``Hierarchical Equations of Motion Approach to Quantum Thermodynamics'', 
in {\it Thermodynamics in the Quantum Regime}
Springer, (2018). 

\bibitem{Strathearn2018}
A. Strathearn, P. Kirton, D. Kilda, J. Keeling, and B.W. Lovett,
``Efficient non-Markovian quantum dynamics using time-evolving matrix product operators'',
{\it Nat. Comm.} \textbf{9}, 3322 (2018).

\bibitem{Nielsen1998} 
M. A. Nielsen, 
PhD Thesis, University of New Mexico, 
quant-ph/0011036  (1998).

\bibitem{Arnesen2001}
M. C. Arnesen, S. Bose, and V. Vedral,
``Natural Thermal and Magnetic Entanglement in the 1D Heisenberg Model'', 
{\it Phys. Rev. Lett.} \textbf{87}, 017901 (2001).

\bibitem{Zhang2005} 
G.-F. Zhang and S.-S. Li,
``Thermal entanglement in a two-qubit Heisenberg XXZ spin chain under an inhomogeneous magnetic field''
{\it Phys. Rev. A} \textbf{72}, 034302 (2005).

\bibitem{Schollwoeck}
U. Schollwoeck,
``The density-matrix renormalization group'',
{\it Rev. Mod. Phys.} \textbf{77}, 259 (2005).

\bibitem{Dwave}
A.D. King, J. Carrasquilla, {\it et al.}, 
``Observation of topological phenomena in a programmable lattice of 1,800 qubits'',
{\it Nature} \textbf{560}, 456 (2018)

\bibitem{Rusconi2017}
C.C. Rusconi, V. Pöchhacker, K. Kustura, J.I. Cirac, and O. Romero-Isart, 
``Quantum Spin Stabilized Magnetic Levitation'',
{\it Phys. Rev. Lett.} \textbf{119}, 167202 (2017)

\bibitem{Pino2018} 
H. Pino, J. Prat-Camps, K. Sinha, B. Prasanna Venkatesh, O. Romero-Isart,  
``On-chip quantum interference of a superconducting microsphere'',
{\it Quantum Sci. Technol.} {\bf 3}, 025001 (2018).




















\end{thebibliography}
\end{document}